\documentclass[a4paper,fleqn]{cas-dc}
\usepackage[switch]{lineno}

\usepackage[authoryear]{natbib}

\def\tsc#1{\csdef{#1}{\textsc{\lowercase{#1}}\xspace}}
\tsc{WGM}
\tsc{QE}

\begin{document}
\let\WriteBookmarks\relax
\def\floatpagepagefraction{1}
\def\textpagefraction{.001}

\shorttitle{Atmospheric dust with OMEGA/MEx}

\shortauthors{Y. Leseigneur \& M. Vincendon}  

\title [mode = title]{OMEGA/Mars Express: a new Martian atmospheric dust hunter}



%

\author[a]{Yann Leseigneur}
\cormark[1]


\ead{yann.leseigneur@ias.u-psud.fr}



\affiliation[a]{organization={Institut d'Astrophysique Spatiale, Université Paris-Saclay, CNRS},
            city={Orsay},
            postcode={91405}, 
            country={France}}

\author[a]{Mathieu Vincendon}









\begin{abstract}
While dust is a key parameter of Mars climate, its behaviour from one year to the next can appear erratic. This variability is notably related to Global Dust Storms (GDS) which occur only certain years with different onset, duration and intensity. The interannual variabilities of the dust cycle may notably explain some characteristics of Recurring Slope Lineae (RSL), slope flows once thought to be caused by liquid water. Long-term monitoring of dust dynamics is thus required to better understand surface-atmosphere dust exchanges on Mars. Here we present a new method to detect atmospheric dust as a function of space and time in the OMEGA Near-InfraRed (NIR) dataset. This dataset covers more than three Martian years; it includes the 2007 GDS which seasonality differs from the preceding (2001) and later (2018) GDS. The method is based on the decrease of the atmospheric optical path caused by dust, that can be measured by OMEGA with the 2~µm CO$_2$ absorption band. This measure is converted to a 0.9~µm NIR dust optical depth using notably comparisons with Mars Exploration Rovers measurements. We derive dust optical depth maps and comment on the variability of the dust seasonal cycle before, during and after the 2007 GDS. We also compare OMEGA NIR optical depths to Thermal InfraRed (TIR) ones derived by other studies. We found a NIR/TIR dust extinction optical depth ratio of 1.8 on average, with some variations notably related to dust particle size. Finally, we show in the northern hemisphere that atmospheric dust and RSL activity is correlated. This may indicate that dust lifting or transport mechanisms working at regional scale also participate to local RSL activity.
\end{abstract}




\maketitle 

\section{Introduction}\label{section_intro}

Dust is omnipresent on Mars, both on the surface and suspended in the atmosphere. This high albedo material modifies solar light transfer and hence energy balance. Therefore, dust is a key parameter of the Martian climate, affecting significantly the thermal response of the surface and the circulation of the atmosphere \citep{smith_interannual_2004, geissler_three_2005,szwast_surface_2006, Kahre2013}. Dust in the atmosphere is composed of mineral particles with a size distribution effective radius typically between 1~µm and 2~µm \citep{Toon1977, Clancy2003, wolff_constraints_2003, Goetz_2005, wolff_constraints_2006, clancy_extension_2010}. Larger values have been reported during 2001 and 2018 global dust storms \citep{wolff_constraints_2003, lemmon_large_2019}. Their composition appears homogeneous over the planet \citep{Bell2000, lasue_comparison_2022} which reflects the global redistribution of dust.

The high mobility of dust leads to movements over different spatial and time scales, from sub-km to the whole planet and from minutes to several years. Dust devils are rapid and local dust movements which occur at the boundary between surface and atmosphere \citep{Balme2003, Cantor2006}. Dust storms occur at larger scales from local (i.e., < 2000 km) to regional extent \citep{martin_analysis_1993}. In a few cases they can reach a planetary scale and form what is named planet-encircling dust events or Global Dust Storms (hereinafter GDS) \citep{cantor_moc_2007,wang_origin_2015}. GDS initiation, propagation and decay can last several months \citep{zurek_interannual_1993}.
 
Most of these movements have a specific seasonality known as the Martian dust cycle. With more than 40 years of orbital observations of Mars, many features of this cycle have been characterised. A Martian Year (MY) can be divided into a "clear season" from a solar longitude ($L_s$: the ecliptic longitude of the sun) of $0^\circ$ to $\sim150^\circ$ associated with an overall lower amount of atmospheric dust, and a "dust season" with a higher dust lifting activity from $L_s\sim150^\circ$ to $\sim360^\circ$ \citep{smith_interannual_2004, montabone_eight-year_2015}. Local and regional dust storms are frequently observed at  most latitudes during the dust season. During the clear season, some storms are still observed but mostly near the northern and southern polar cap edges \citep{battalio_mars_2021,sanchez-lavega_cellular_2022}. Some dust storms flush from one hemisphere to the other through dust storm travel routes, mainly from the northern hemisphere (Acidalia-Chryse $\sim 325^\circ$E and Utopia-Isidis $\sim 90^\circ$E routes, \cite{wang_origin_2015,battalio_mars_2021}), but also from the southern hemisphere (Aonia/Solis-Valles Marineris $\sim 260-320^\circ$E route, \cite{battalio_aonia-solis-valles_2019}).

Even if several features of this seasonal dust cycle reliably repeat from one year to the next, there are still some interannual variations. The most striking change is the irregular occurrence of GDS certain Martian years. GDS can have different origins: some appear during the early dust season (near $L_s \sim 180^\circ$) such as the 2001 (MY~25) GDS \citep{strausberg_observations_2005,cantor_moc_2007} and the 2018 (MY~34) GDS \citep{guzewich_mars_2019,lemmon_large_2019, montabone_martian_2020}, while others onset in the middle of the dust season near $L_s \sim 260^\circ$ (e.g., the 2007/MY~28 GDS, see \cite{lemmon_dust_2015} and \cite{wang_origin_2015}). These storms influence the atmospheric dynamics during the event \citep{bertrand_simulation_2020,montabone_martian_2020}. They are also responsible for the main modifications of the surface dust cover \citep{geissler_three_2005, szwast_surface_2006, vincendon_mars_2015}. However, all the formation mechanisms and consequences of GDS are not yet known precisely \citep{kahre_simulating_2005,newman_impact_2015,bertrand_simulation_2020}. We can for example wonder whether precursory signs herald GDS the season or the year before. Symmetrically, we could ask if a GDS leads to a specific dust cycle the year after due to major modifications of dust reservoirs. The impact of a GDS may also differ depending on its onset timing (early or middle of the dust season) and duration. Thus, long-term monitoring of dust dynamics may help answering these questions.

Another open question regarding the dust cycle is its precise link with Recurring Slope Lineae (hereinafter RSL). RSL, discovered by \cite{mcewen_seasonal_2011}, are seasonal dark flows on Martian steep slopes. These features have activity phases when they appear and grow downwards more or less incrementally, and have disappearance phases when they fade (partially or totally) more or less progressively. While several studies were first conducted to investigate a possible wet origin for RSL (involving liquid water, brines, etc., see for example a review in \cite{stillman_chapter_2018}), they are now principally explained by dry processes involving dust \citep{schmidt_formation_2017,dundas_granular_2017,schaefer_case_2019,vincendon_observational_2019,dundas_aeolian_2020,mcewen_mars_2021,munaretto_multiband_2022}. While their formation mechanisms are not yet precisely understood, it has been suggested that RSL could primarily be dust-removed features \citep{vincendon_observational_2019, mcewen_mars_2021}, and that the timing of local RSL activity may be related to that of the global seasonal dust cycle \citep{vincendon_observational_2019}. Actually, major RSL formation events occur in southern summer for all latitudes \citep{mcewen_seasonal_2011, stillman_chapter_2018, vincendon_observational_2019}, which is also the season of main dust storms on Mars. Additionally, RSL are more abundant after a GDS decay \citep{mcewen_mars_2021}, and some RSL sites were observed to be active only after GDS \citep{vincendon_observational_2019, mcewen_mars_2021}. In more details, RSL formation and lengthening timing however differ between northern and southern hemispheres, and between some RSL sites of a given hemisphere \citep{mcewen_seasonal_2011,McEwen2014,stillman_observations_2016, stillman_two_2018}. A more precise study of the connections between local/regional/global atmospheric dust activity, and local RSL activity, could bring additional clues to understand precisely the RSL formation mechanisms.

Several instrumental methods can be used to study dust on Mars. The dust storms are observed from Earth ground observations as far back as the early nineteenth century (observations of "yellow clouds") and were first associated with dust storms by \cite{antoniadi_planete_1930}. Since the 1970's, different instruments were sent to Mars onboard different spacecrafts (orbiters, landers and rovers) providing key features on the dust cycle (see an overview in \cite {smith_spacecraft_2008}). Visible cameras are useful from the orbit to identify dust storms and to follow their growth and their movements with daily global imaging of Mars \citep{cantor_martian_2001, cantor_moc_2007, battalio_mars_2021}. Such visible cameras are also used on rovers to measure the sun extinction and thus the atmospheric extinction optical depth as for the Mars Exploration Rovers (MER: Spirit and Opportunity) \citep{lemmon_dust_2015}. Thermal-infrared spectrometers have been widely used from the orbit to retrieve column dust optical depths from temperature profiles (mostly at 9~µm) with nadir-observations \citep{smith_interannual_2004, montabone_eight-year_2015}. Orbital observations of the limb with various wavelength ranges are another powerful way to constrain dust vertical distribution \citep{rannou_dust_2006, Smith2013, daversa_vertical_2022}.

Two near-InfraRed (NIR hereinafter) imaging spectrometers are also present on Mars orbiters: OMEGA and CRISM which are respectively onboard Mars Express and Mars Reconnaissance Orbiter. While their main purposes are to study the surface, these spectrometers are also useful to constrain atmospheric properties as they enable composition identification \citep{Encrenaz2005, vincendon_new_2011, madeleine_aphelion_2012, Clancy2012}. OMEGA NIR channel (named C-channel) operated over more than 3 Martian years, making it possible to explore seasonal and interannual variations of atmospheric phenomena over the whole planet. Some OMEGA studies dedicated to atmospheric dust have already been done such as: calculation of the optical depth at polar latitudes \citep{Vincendon2007, Vincendon2008}, local dust storms characterisation \citep{maattanen_study_2009, Oliva2018} and vertical dust retrievals from limb observations \citep{daversa_vertical_2022}. However, global monitoring of the atmospheric dust has not yet been carried out.

In this paper, we develop and apply a new method to calculate atmospheric dust optical depth with the OMEGA dataset. We first present OMEGA data and the filtering criteria used in this study in section~\ref{section_data}. Then, we detail the method to detect atmospheric dust and compute dust optical depth in section~\ref{section_method}. Finally, we analyse in section~\ref{section_results&discussions} the time and spatial variations of atmospheric dust over more than 3 Martian years including the 2007 (MY~28) GDS. Summary and conclusions are drawn in section~\ref{section_conclusion}.

\section{Data}\label{section_data}

\subsection{OMEGA}\label{subsection_OMEGA}
We use the visible and NIR imaging spectrometer OMEGA on board Mars Express in orbit since late December 2003 (since 19 years!). OMEGA observes both the surface and the atmosphere with its nominal nadir viewing pointing geometry. The dataset is composed of hyperspectral observations with an average spatial resolution of 1 km (it varies between 0.3 and $\sim$ 4 km depending of the Mars Express position on its elliptical orbit around Mars). The spectral range, from 0.38 to 5.1~µm, is covered by three channels (called “V”, “C” and “L” channels) as defined in \cite{2004ESASP1240...37B}.
For this study, we mainly used the C-channel (0.93-2.73~µm) which has been operating from 2004 to 2010\footnote{We can notice that V-channel and L-channel are still operating at the time of writing of this article.}, i.e. over more than three Martian Years (MY) notably one year before (MY~27), during (MY~28) and after (MY~29) a Global Dust Storm (hereinafter GDS). This dataset contains 9646 observations and covers almost the entire Martian surface with repeated observations available for most places. The L-channel (2.53-5.1~µm) is also used to compute spectral indexes defined in the following section.

OMEGA data are handled using the Python module “OMEGA-Py” freely available on GitHub at \url{https://github.com/AStcherbinine/omegapy}.

As detailed in the next sections, our study relies on the quantification of the variability of the atmospheric 2~µm carbon dioxide (CO$_2$) absorption as a function of atmospheric dust content. This band has been already used by \cite{forget_remote_2007} and \cite{spiga_remote_2007} to compute the pressure at the surface from the Martian orbit.

\subsection{Data filtering}\label{subsection_filters}

\begin{table*}[width=\textwidth,cols=4, pos=h]
\caption{Description of data filtering criteria. $R$ is the reflectance factor: $I/(F \cos(i))$. Data quality is an index (integer between 1 and 5, respectively lowest and highest quality, see OMEGA software "SOFT10" documentation for more details) that evaluates the quality of the OMEGA observations. (Note that before the orbit 1900, the data quality value 3 corresponds to 5 for the rest of the dataset.)} \label{table_filtres}
\begin{tabular*}{\tblwidth}{@{}LL@{}CC@{}@{}CC@{}CC@{}}
\toprule
  & Description & Formulation & Detection & Comments\\ 
  & & &  threshold &\\ 
\midrule
 & H$_2$O$(1.5\ \mbox{µm})$ & $1 - \mbox{R}(1.500 \ \mbox{µm}) / (\mbox{R}(1.385 \ \mbox{µm})^{0.7}$ & 0.01 & Values > 0.01 present water ice.\\
 & & $\times \ \mbox{R}(1.772 \ \mbox{µm})^{0.3})$ & \\
 & Ice clouds index & $\mbox{R}(3.400 \ \mbox{µm}) / \mbox{R}(3.525 \ \mbox{µm})$ & 0.70 & Values < 0.70 present water ice clouds.\\
 & Polar cap edge & \cite{kieffer_mars_2000,kieffer_tes_2001}; & / & $3^\circ$ latitude margin is added.\\
 & & \cite{titus_mars_2005} \\
 & Incidence ($i$) & Data & $80^\circ$ & Pixels with $i<80^\circ$ are considered.\\
 & Emergence ($e$) & Data & $10^\circ$ & Pixels with $e<10^\circ$ are considered.\\
 & Data quality & Data & $4$ & $\mbox{Data quality} \geq 4$ are considered.\\ 
\bottomrule
\end{tabular*}
\end{table*}

We have applied different filters to select observations relevant for our study and to remove artefacts. The definitions and thresholds of these filters are detailed in table~\ref{table_filtres}. Most criteria are routinely used while handling the OMEGA dataset, and have been previously described \citep{ody_global_2012,vincendon_mars_2015}.

We pay here a particular attention to data properties that may impact the 2~µm range. In particular, transient H$_2$O or  CO$_2$ ice at the surface or in the atmosphere that can create a spectral signature at 2~µm that may modify the quantification of the CO$_2$ gas band depth. Firstly, we remove observations above or near the seasonal polar caps. The position  of ice cap edges changes with season: we evaluate ice spatial extent using thermal infrared data of the northern \citep{kieffer_tes_2001} and southern hemisphere \citep{kieffer_mars_2000,titus_mars_2005}. We exclude areas corresponding to the caps, with a 3$^\circ$ equatorward latitude margin. Additionally, we identify water ice (clouds or frost) based on the 1.5~µm water ice band \citep{langevin_observations_2007}. A conservative threshold of 1\% is considered (this may remove observations that actually do not contain ice). An additional water ice criteria based on the 3~µm band is also used (Table~\ref{table_filtres}) as it can help to detect some clouds \citep{langevin_observations_2007,madeleine_aphelion_2012,audouard_water_2014}.

Then we filter data depending on their viewing geometry. We consider only near nadir observations (emergence angle $e<10^\circ$) with an incidence angle ($i$) lower than 80$^\circ$ to simplify optical path modelling in the next sections. As most OMEGA observations is compliant with these restrictions, this only weakly impact the amount of usable data.

\section{Method}\label{section_method}

\subsection{Overview}\label{subsection_principle}
\begin{figure*}[!h]
	\centering
		\includegraphics[width=0.8\textwidth]{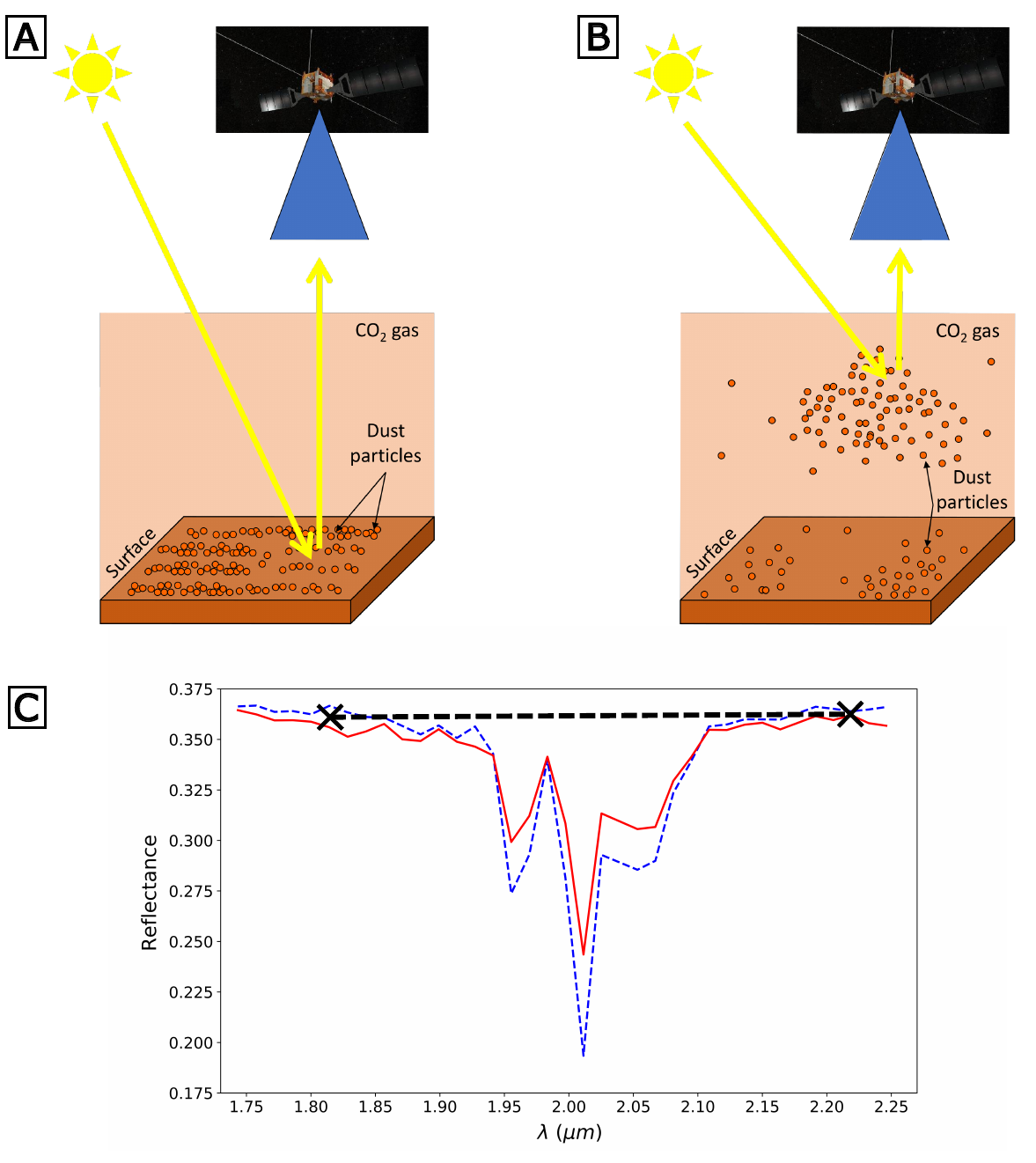}
	  \caption{Overview of the method used in this paper to detect dust in the atmosphere. (Top) schemes illustrating photon paths in the atmosphere in case of low [A] and high [B] atmospheric dust load. [C] represents the corresponding 2~µm CO$_2$ absorption bands (clear atmosphere: dashed blue line; dusty atmosphere: red line). The band is shallower in a dusty atmosphere than in a clear one. The (linear) continuum used for band depth calculation is represented by a black dotted line: it is defined between wavelengths 1.82~µm (average of 1.80~µm and 1.84~µm) and 2.22~µm symbolised by black crosses.}\label{graph_method}
\end{figure*}

We have developed a method to identify atmospheric dust based on the 2~µm CO$_2$ gas absorption band. The depth of this band depends on the distance travelled by the sun’s rays through the atmosphere mainly composed of this gas ($\sim$~96\%). Indeed, in a clear atmosphere, with a low atmospheric dust load, the sun’s rays reach the surface where they are reflected towards the instrument (Figure~\ref{graph_method}, panel~A). As represented in panel~C of Figure~\ref{graph_method}, the 2~µm band corresponding to this situation (in blue) is deep. However, when there is a high atmospheric dust load, the sun’s rays are reflected on the atmospheric dust. Thus, the distance travelled in CO$_2$ is shorter, as we can see in panel~B of  Figure~\ref{graph_method}, and the corresponding band is therefore less deep (red band in panel~C). Note that a method presenting some similarities has been previously suggested with the 2.7~µm CO$_2$ band, which differs from the 2~µm one as it saturates to 0 without atmospheric dust \citep{titov_new_2000}.

In the next sections, we quantify this modification of the CO$_2$ optical depth by comparing the observed optical depth ($\tau_{CO_{2},\ obs}$) of a given OMEGA observation to the theoretical one for a clear atmosphere ($\tau_{CO_{2}, \ pred}$). The resulting difference is $\Delta \tau_{CO_{2}}$ (equation \ref{eq_delta_tau}).

\begin{equation}
    \Delta \tau_{CO_{2}} = \tau_{CO_{2},\ pred} - \tau_{CO_{2},\ obs}
    \label{eq_delta_tau}
\end{equation}

 This quantity will then be related to the optical depth of atmospheric dust $\tau_{dust}$ using a calibration procedure based on ground truth measurements provided by the Mars Exploration Rovers (equation \ref{eq_delta_tau_mer}).
 
 \begin{equation}
    \tau_{dust} = f(\Delta \tau_{CO_{2}})
    \label{eq_delta_tau_mer}
\end{equation}

\subsection{Calculation of CO$_2$ optical depths}\label{subsection_model}
We compute the observed CO$_2$ optical depth for a given OMEGA pixel at 2~µm ($\tau_{CO_{2},\ obs}$) using equation~\ref{eq_Beer_Lambert}, where $R(2.01 ~ \mbox{µm})$ is the reflectance factor at 2.01~µm and $R_0$ is the reflectance factor of the continuum (see panel~C of Figure~\ref{graph_method} for details).

\begin{equation}
    \frac{R(2.01 \ \mbox{µm})}{R_0} = e^{-\tau_{CO_{2},\ obs}}
    \label{eq_Beer_Lambert}
\end{equation}

We estimate the predicted CO$_2$ optical depth when the atmosphere is clear ($\tau_{CO_{2},\ pred}$) using the following approach. The optical depth depends on the amount of gas encounter by the photons as they pass through the atmosphere. Firstly, in the simple case of normal incidence rays ($i=0^\circ$), the CO$_2$ optical depth is proportional to the pressure $P$ (equation~\ref{tau_pred_basic}). We note $\alpha$ the proportionality factor which depends notably on OMEGA instrumental response and cannot be simply inferred.

\begin{equation}
    \tau_{CO_{2},\ pred} = \alpha \ P
    \label{tau_pred_basic}
\end{equation}

For other incidence values, the geometrical path of the sun's rays should be $1+1/\cos(i)$. However, a small amount of dust is always present in the atmosphere of Mars, even during the clear atmospheric season. As a consequence, the actual optical path will slightly depart from this law. We take this into account by adding an exponent $d <1$ to the term $1/\cos(i)$ (see equation~\ref{eq_tau_pred_expo}).

\begin{equation}
    \tau_{CO_{2},\ pred} = \alpha \ P \times \Big (1 + \Big (\frac{1}{\cos(i)}\Big )^d \Big )
    \label{eq_tau_pred_expo}
\end{equation}

The presence of residual atmospheric dust when the atmosphere is clear will also result in the fact that $\tau_{CO_{2},\ pred}$ is a function of surface albedo. Indeed, the average atmospheric optical path measured from the orbit depends on the relative weight of photons reaching the surface and then scattered (long path), versus photons directly scattered by dust before reaching the surface (shorter path). In the following "albedo" will refer to the reflectance factor at 2.26~µm (this wavelength is located outside the 2~µm band, it is used to estimate the reflectance level of the surface at 2 µm). Overall, $\tau_{CO_{2},\ pred}$ is thus expected to be describable by equation~\ref{equation_model}, where $f(\mbox{Alb})$ is an unknown albedo function.

\begin{equation}
    \tau_{CO_{2},\ pred} = \alpha \ P \times \Big (1 + \Big (\frac{1}{\cos(i)}\Big )^d \Big ) \times f(\mbox{Alb})
    \label{equation_model}
\end{equation}

In the next section, we calibrate the unknown parameters $\alpha$, $d$ and $f(\mbox{Alb})$ of equation~\ref{equation_model}.

\subsection{Calibration of clear atmosphere CO$_2$ optical depth}\label{subsection_model_calibration}

We made a selection of 57 OMEGA observations (see Table~\ref{table_obs_calib}) to calibrate equation~\ref{equation_model}. As $\tau_{CO_{2},\ pred}$ corresponds to clear atmospheric conditions, we have selected data corresponding to expected low optical depths according to previous studies \citep{smith_interannual_2004,lemmon_dust_2015}. These selected observations cover various altitudes and seasons, sampling all expected Martian pressures. Similarly, we have selected observations at various solar incidence angles and surface albedos, useful to characterise these two other dependencies of $\tau_{CO_{2},\ pred}$.

\begin{table*}[width=\textwidth,cols=5, pos=h]
\caption{OMEGA calibration data used to characterise the three CO$_2$ optical depth dependencies with the corresponding range widths for each one ($\forall$: all values are considered).}\label{table_obs_calib}
\begin{tabular*}{\tblwidth}{@{}CC@{}l@{}CC@{}CC@{}CC@{} }
\toprule
  & Dependency & Cubes & Pressure range & Albedo range & Incidence range\\ 
\midrule
 & Albedo & 2133$\_$5, 5396$\_$2, 5450$\_$2 & $\Delta P \leq 10 \ \mbox{Pa}$ & $\forall$ & $\Delta i \leq 3^\circ$\\
 & Incidence & 0049$\_$0, 0097$\_$0, 0103$\_$3, 0243$\_$3, 0248$\_$2, & $\Delta P \leq 10 \ \mbox{Pa}$ & $\Delta \mbox{Alb} \leq 0.10$ & $\forall$ \\
 & & 0308$\_$2, 1170$\_$3, 1533$\_$2, 2432$\_$0, 2436$\_$1, & & & &\\
 & & 2526$\_$1, 3639$\_$2, 5398$\_$2, 5450$\_$2, 6486$\_$5, & & & &\\
 & & 7267$\_$3, 7287$\_$3, 7287$\_$4, 7288$\_$2, 7290$\_$3, & & & &\\
 & & 7290$\_$4, 7294$\_$3, 7303$\_$3, 7306$\_$4, 7310$\_$2, & & & &\\
 & & 7324$\_$3, 7327$\_$3, 7330$\_$3, 7334$\_$3, 7365$\_$3, & & & &\\
 & & 7366$\_$4, 7370$\_$3, 7393$\_$4, 7422$\_$3 & & & &\\
 & Pressure & 0049$\_$0, 0097$\_$0, 0103$\_$3, 0243$\_$3, 0248$\_$2, & $\forall$ & $\forall$ & $\forall$\\
 & & 0270$\_$3, 0286$\_$3, 0308$\_$2, 1143$\_$3, 1150$\_$3, & & & &\\
 & & 1170$\_$3, 1183$\_$3, 1238$\_$2, 1510$\_$5, 1533$\_$2, & & & &\\
 & & 2133$\_$5, 2245$\_$4, 2256$\_$4, 2432$\_$0, 2436$\_$1, & & & &\\
 & & 2526$\_$1, 3198$\_$3, 3220$\_$1, 3262$\_$2, 3345$\_$2, & & & &\\
 & & 3639$\_$2, 3674$\_$2, 3685$\_$2, 5396$\_$2, 5398$\_$2, & & & &\\
 & & 5450$\_$2, 5692$\_$2, 5814$\_$2, 6486$\_$5, 7267$\_$3, & & & &\\
 & & 7287$\_$3, 7287$\_$4, 7288$\_$2, 7290$\_$3, 7290$\_$4, & & & &\\
 & & 7294$\_$3, 7303$\_$3, 7306$\_$4, 7310$\_$2, 7324$\_$3, & & & &\\
 & & 7327$\_$3, 7330$\_$3, 7333$\_$3, 7334$\_$3, 7337$\_$3, & & & &\\
 & & 7365$\_$3, 7366$\_$4, 7370$\_$3, 7392$\_$4, 7393$\_$4, & & & &\\
 & & 7422$\_$3, 7907$\_$3 & & & &\\
\bottomrule
\end{tabular*}
\end{table*}

\begin{figure}[!h]
	\centering
	\includegraphics[width=0.5\textwidth]{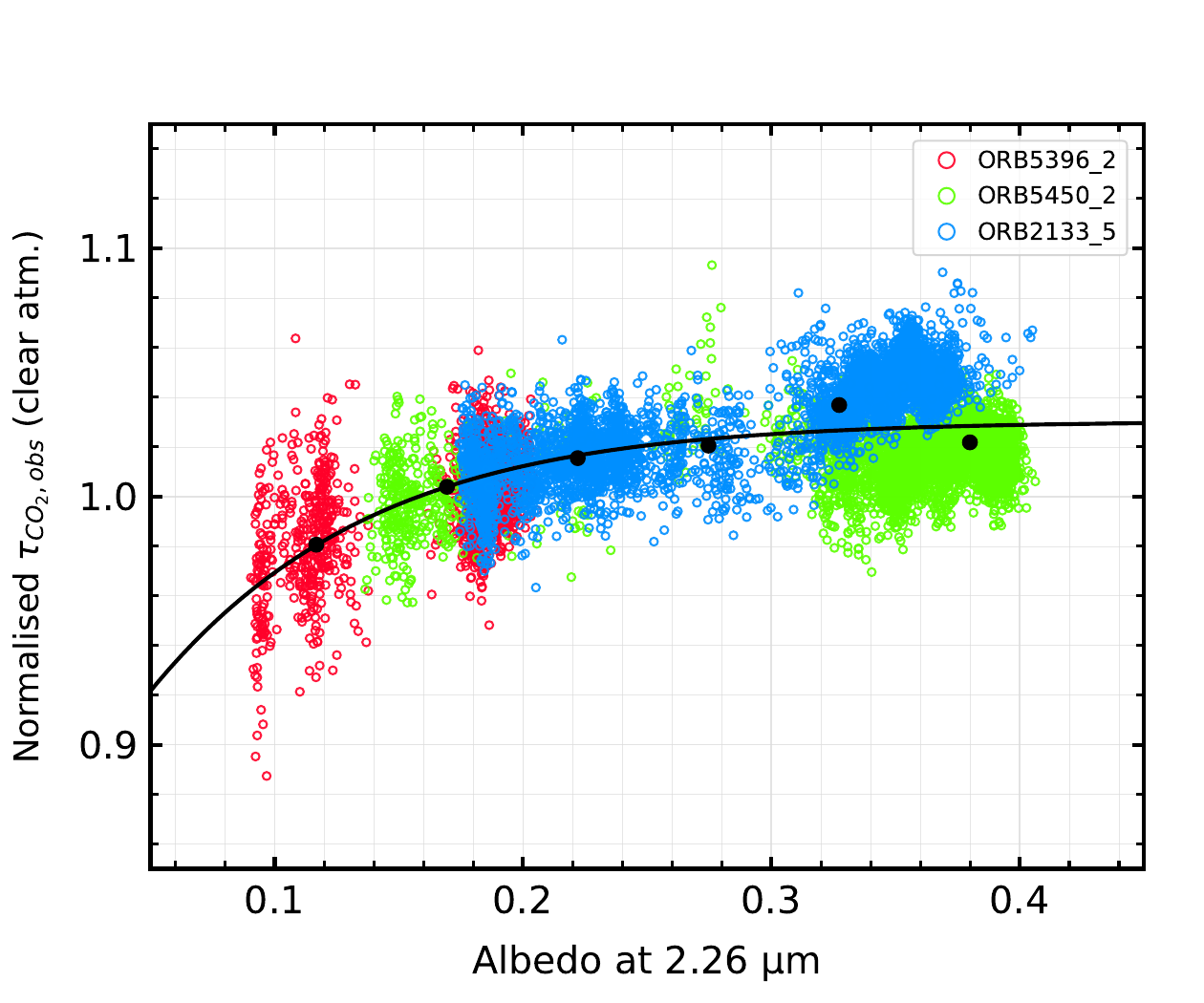}
	 \caption{Calibration of the albedo dependence $f(\mbox{Alb})$ of $\tau_{CO_{2},\ pred}$ (equation~\ref{equation_model}). $f(\mbox{Alb})$ (solid black line, equation~\ref{eq_albedo_calib}) is constrained using observed variations of clear atmosphere CO$_2$ optical depth (normalised) as a function of albedo, derived from 3 OMEGA observations (colors dots). Other variables (incidence and pressure) are constant in each observation. Median values are computed on albedo ranges of 0.06 width (black circles) to better constrain the fit.}
	  \label{graph_albedo_calib}
\end{figure}

Firstly, we characterised the albedo dependency of $\tau_{CO_{2},\ pred}$ using three observations for which albedo varies strongly while other variables (pressure and incidence) are nearly constant (variations respectively less than $10~\mbox{Pa}$ and less than $3^\circ$, see Table~\ref{table_obs_calib}). We show in Figure~\ref{graph_albedo_calib} how the observed CO$_2$ optical depth changes with albedo. We observe that variations are lower than $\pm~5 \%$. We use these observations to estimate the expression of $f(\mbox{Alb})$ (equation~\ref{eq_albedo_calib}) and its parameter values (see Table~\ref{table_values_param}).

\begin{equation}
    f(\mbox{Alb}) = \frac{A_{max}}{1 + \Big(\frac{A_{max}}{A_0}-1\Big) e^{{-k \times \mbox{Alb}}}}
    \label{eq_albedo_calib}
\end{equation}

\begin{figure}[!h]
	\centering
		\includegraphics[width=0.5\textwidth]{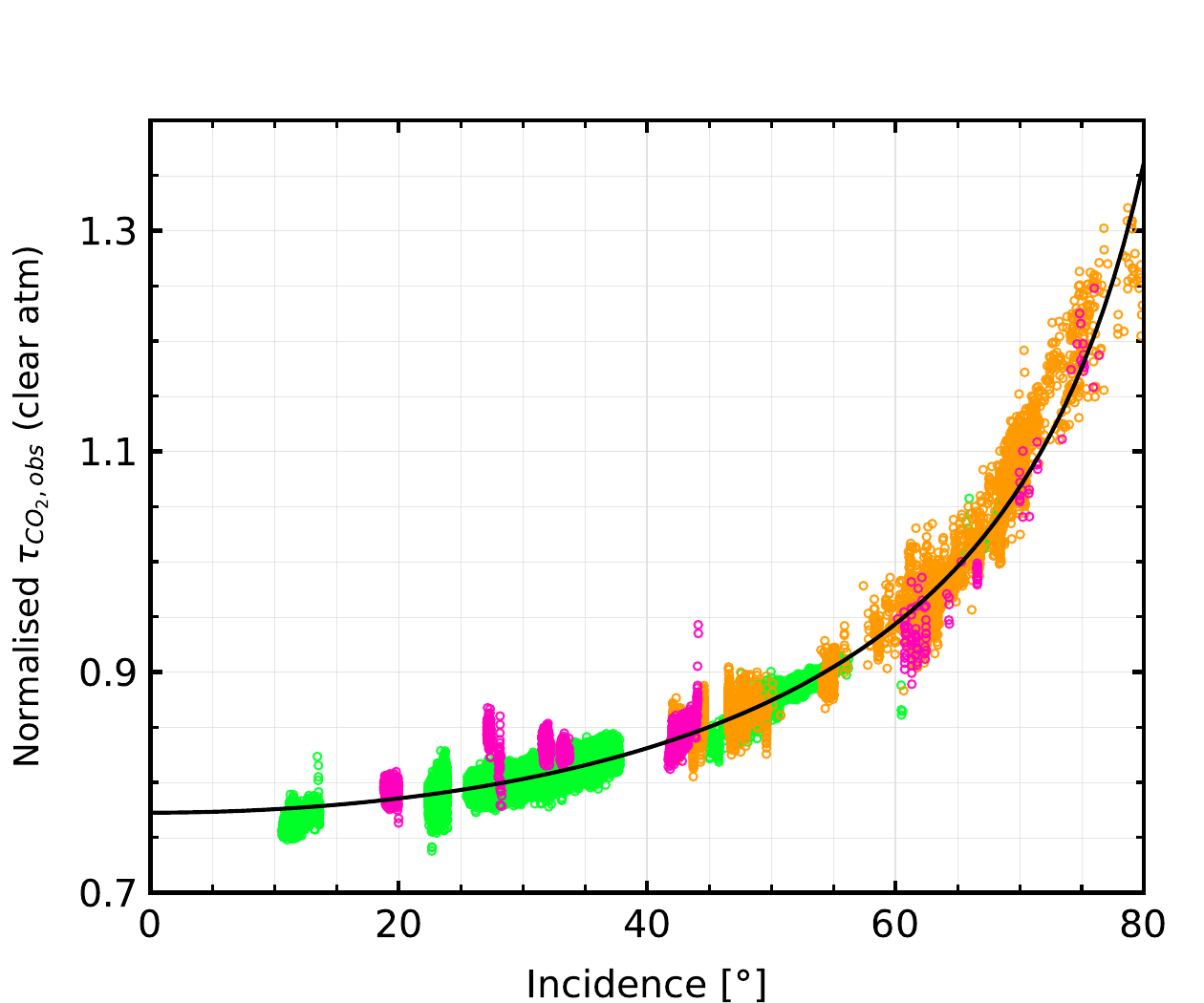}
	  \caption{Calibration of the incidence dependency of $\tau_{CO_{2},\ pred}$ (equation~\ref{equation_model}). This graph represents the observed CO$_2$ optical depth (normalised) as a function of incidence. 34 OMEGA observations for three different pressure and albedo ranges are used, respectively: $[500, \ 510]~\mbox{Pa}$ and $[0.35, \ 0.45]$ (pink points), $[720, \ 730]~\mbox{Pa}$ and $[0.35, \ 0.45]$ (green points), $[485, \ 495]~\mbox{Pa}$ and $[0.25, \ 0.30]$ (orange points). An incidence exponent $d \simeq 0.53$ in equation~\ref{equation_model} provides a satisfactory fit with the data (solid black line).}\label{graph_inci_calib}
\end{figure}

\begin{figure}[!h]
	\centering
		\includegraphics[width=0.5\textwidth]{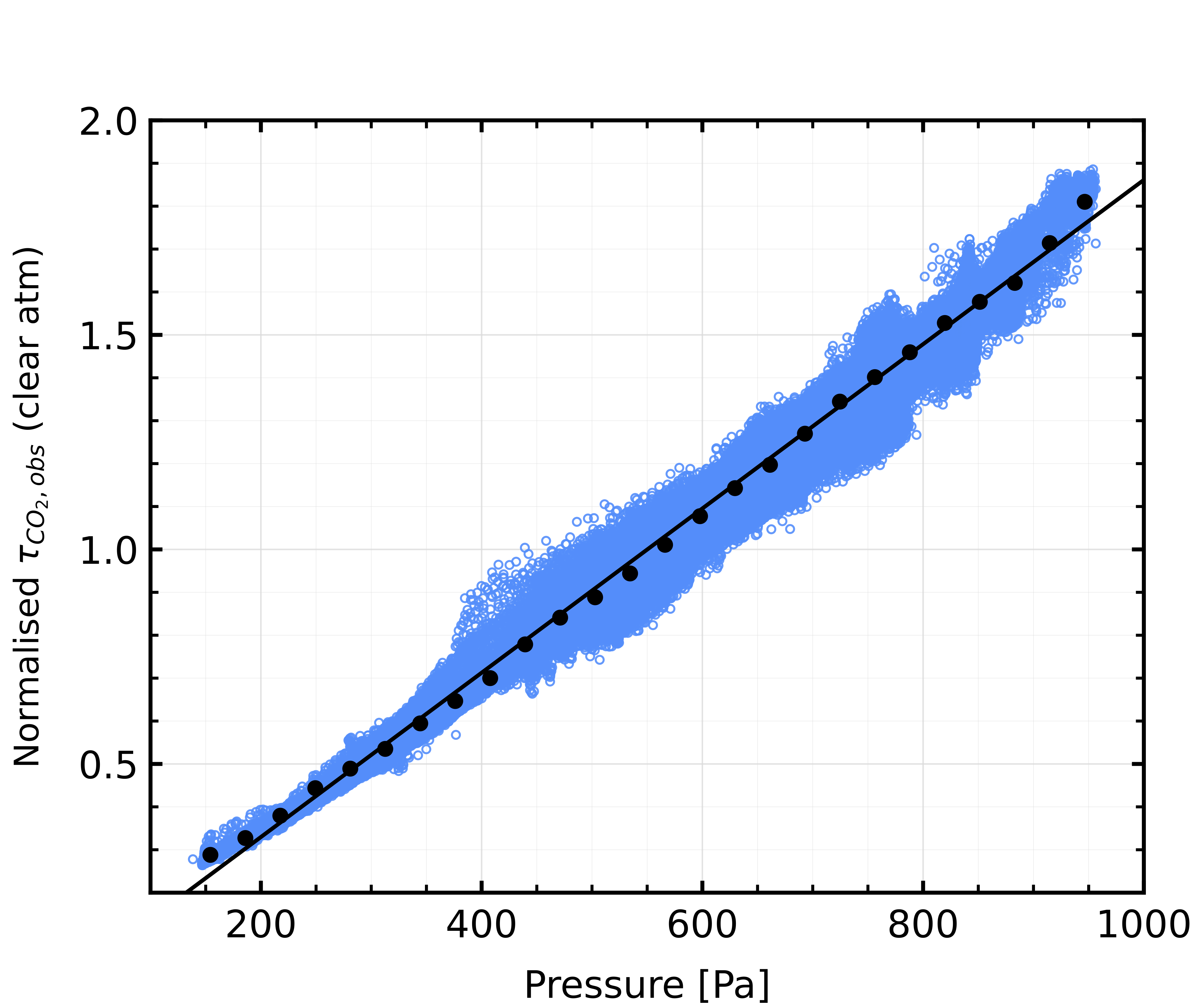}
	  \caption{Calibration of $\tau_{CO_{2},\ pred}$ pressure dependency (equation~\ref{equation_model}). This graph represents the observed CO$_2$ optical depths of 57 OMEGA observations (blue circles) as a function of pressure. CO$_2$ optical depths are normalised to take into account albedo and incidence variations previously determined in  Figures~\ref{graph_albedo_calib} and \ref{graph_inci_calib}. Median values are computed on pressure ranges of $32~\mbox{Pa}$ width (black circles) to better constrain the linear fit (solid black line; see Table~\ref{table_values_param} for parameter values).}\label{graph_pressure_calib}
\end{figure}

Secondly, we use 34 observations covering a large incidence range to estimate the value of the exponent $d$ (equation~\ref{equation_model}). Other variables (pressure and albedo) are nearly constant (variations respectively less than $10~\mbox{Pa}$ and less than $0.10$, see Table~\ref{table_obs_calib} and Figure~\ref{graph_inci_calib} for more details). The behaviour of the observed CO$_2$ optical depth as a function of incidence is shown in Figure~\ref{graph_inci_calib}. We observe that variations are small (about $7\%$) for low-medium incidences ($i<40^\circ$) and large (about $40\%$) for high incidences ($i>60^\circ$). These observations lead to an exponent $d \simeq 0.53$.

Finally, we use 57 observations (see Table~\ref{table_obs_calib}) to constrain the parameter $\alpha$ which characterises variations with pressure (Figure~\ref{graph_pressure_calib}). This last calibration step is done without any restrictions on these observations (albedo nor incidence) as we account for albedo and incidence variations using $f(\mbox{Alb})$ and $d$ described above. Pressure is evaluated using the pressure calculator (pres0) of the Mars Climate Database (MCD, version 4.1) \citep{forget_co2snowfall_1998,millour_mars_2018}, which takes OMEGA pixel information (latitude, longitude, altitude, solar longitude and local time) as inputs. The main hypothesis of the calculator is to consider the total pressure at the surface equivalent to the hydrostatic pressure (the weight of the column of atmosphere above the surface), which is true in the great majority of Martian meteorological conditions. It is also connected to the simulation outputs of the Mars Planetary Climate Model of the \emph{Laboratoire de Météorologie Dynamique} (LMD). Therefore, it takes into account the seasonal variations of the pressure that are linked to the CO$_2$ cycle \citep{forget_co2snowfall_1998}. As expected, we observe in Figure~\ref{graph_pressure_calib} an increasing linear trend between $\tau_{CO_{2},\ pred}$ and pressure. Note however that the intercept is not exactly null. 

Overall, we thus obtain equation~\ref{eq_tau_pred}: associated parameters are summarised in Table~\ref{table_values_param}.

\begin{equation}
    \tau_{CO_{2},\ pred} = \alpha \ P \Big(1 + \Big( \frac{1}{\cos(i)}\Big)^{d}\Big) \times \frac{A_{max}}{1 + \Big(\frac{A_{max}}{A_0}-1\Big) e^{{-k \times Alb}}} \ +b
    \label{eq_tau_pred}
\end{equation}

\begin{table}[!h]
\caption{Parameters values for equations~\ref{equation_model}, \ref{eq_albedo_calib} and \ref{eq_tau_pred}.}\label{table_values_param}
\begin{tabular*}{\tblwidth}{@{}CC@{}CC@{} }
\toprule
  & Parameters & Values\\ 
\midrule
 & $\alpha$ & $7.39 \times 10^{-4}$\\
 & $d$ & $5.29 \times 10^{-1}$\\
 & $A_{max}$ & $1.03$\\
 & $A_0$ & $8.44 \times 10^{-1}$\\
 & $k$ & $1.26 \times 10^{1}$\\
 & $b$ & $-2.39 \times 10^{-2}$\\
\bottomrule
\end{tabular*}
\end{table}

\subsection{Dust optical depth calibration using Mars Exploration Rovers}\label{subsection_MER_comparison}
In this section, we describe how $\Delta \tau_{CO_{2}}$ (equation~\ref{eq_delta_tau}) relates to the dust optical depth ($\tau_{dust}$) using observations gathered in the vicinity of Mars Exploration Rovers (MER hereinafter). These two rovers, Spirit and Opportunity, have operated respectively on Gusev crater and Meridiani Planum, during the same period as OMEGA, between January 2004 and March 2010 (MY~26-29) for Spirit, between January 2004 and February 2019 for Opportunity (MY~26-34). Atmospheric extinction optical depth has been computed in \cite{lemmon_dust_2015} by using the Pancam solar images (0.88~µm) of the two rovers, allowing an accurate characterisation of the annual and interannual dust variability from the Martian surface. We consider that these MER measurements of optical depth ($\tau_{MER}$) are a "ground truth" reference for our calibration procedure. Note that \cite{lemmon_dust_2015} suggest to use these 0.88~µm measurements to compare them to other datasets.

To carry out the correspondence between $\Delta \tau_{CO_{2}}$ and $\tau_{MER}$, we used 38 OMEGA observations near MER landing sites. Two areas of interest have been defined: $[173, \ 177]^\circ$E and $[12, \ 17]^\circ$S for Spirit, $[353.5, \ 355.5]^\circ$E and $[1.7, \ 2.7]^\circ$S for Opportunity. Within these areas we use only OMEGA pixels with altitudes in the $[-2, \ -1.9] \ \mbox{km}$ range for Spirit and $[-1.58, \ -1.38] \ \mbox{km}$ for Opportunity (values close to the rover location altitudes). Note that we apply different spectral filters on OMEGA observations (see section~\ref{subsection_filters}) to exclude, among other things, ice clouds (main aerosols with dust) that can change the atmospheric extinction optical depth. So, it is assumed that for OMEGA/MER co-observations the MER measurements are equivalent to dust extinction optical depths.

\begin{figure}[!h]
	\centering
		\includegraphics[width=0.5\textwidth]{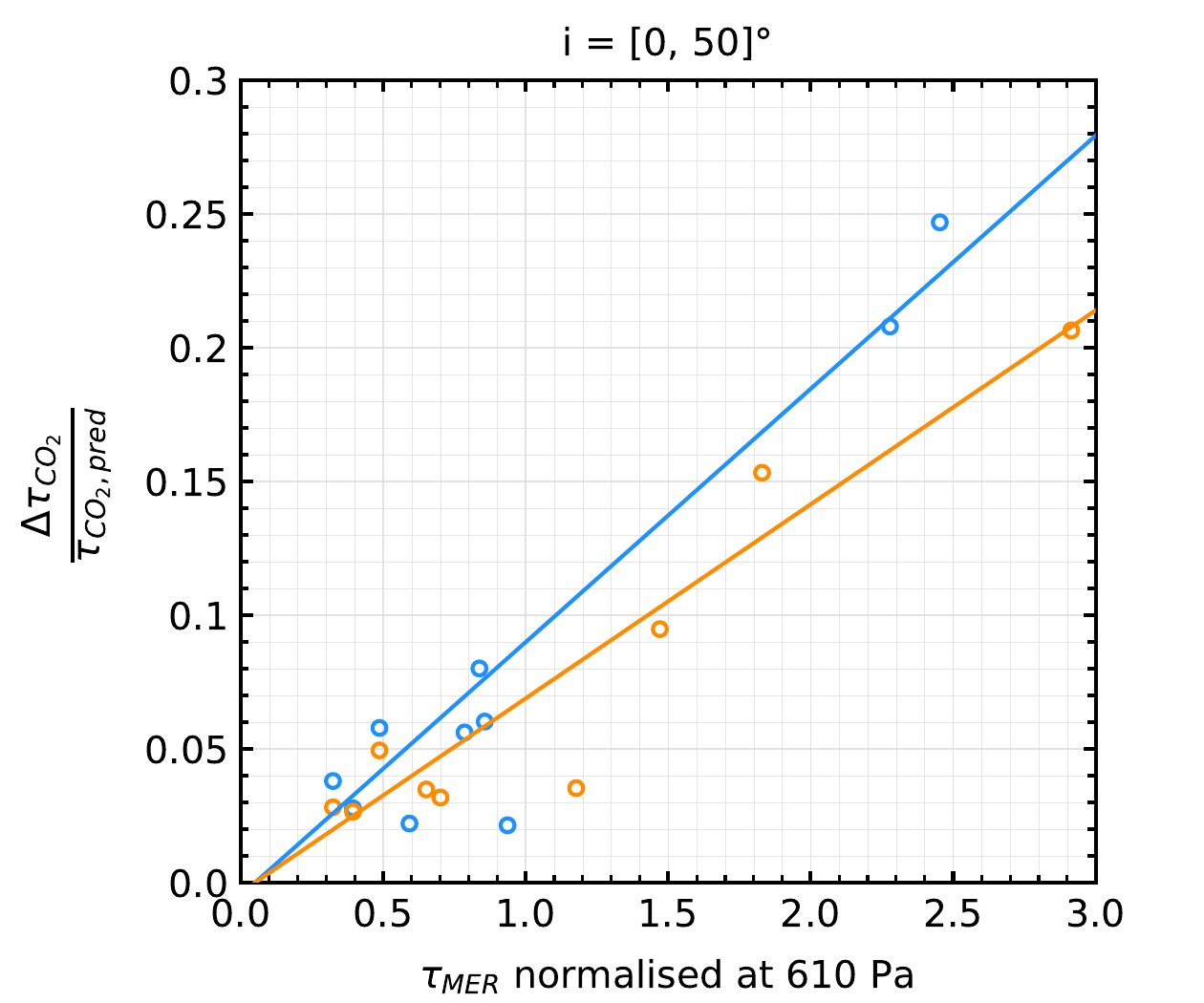}
	  \caption{Relation between $\Delta \tau_{CO_{2}} / \tau_{CO_{2},\ pred}$ and $\tau_{MER}$ for OMEGA observations with $i \leq 50^\circ$. This graph represents observation median values of $\Delta \tau_{CO_{2}} / \tau_{CO_{2},\ pred}$ for two albedo ranges $[0.255, \ 0.265]$ (blue points) and $[0.395, \ 0.405]$ (orange points) as a function of $\tau_{MER}$ normalised at $610~\mbox{Pa}$. A linear fit has been realised for each albedo range (blue and orange solid lines).}\label{graph_delta_tau_tau_MER_linfitAlb_i0-50}
\end{figure}

The quantity that we compare to $\tau_{MER}$ is $\Delta \tau_{CO_{2}} / \tau_{CO_{2},\ pred}$, which is $\Delta \tau_{CO_{2}}$ normalised by the clear atmosphere CO$_{2}$ optical depth. Therefore, our dust index is normalised in pressure. As a consequence, we will compare this index to $\tau_{MER}$ normalised to a constant reference pressure (610 Pa). 

\begin{figure}[!h] 
	\centering
		\includegraphics[width=0.5\textwidth]{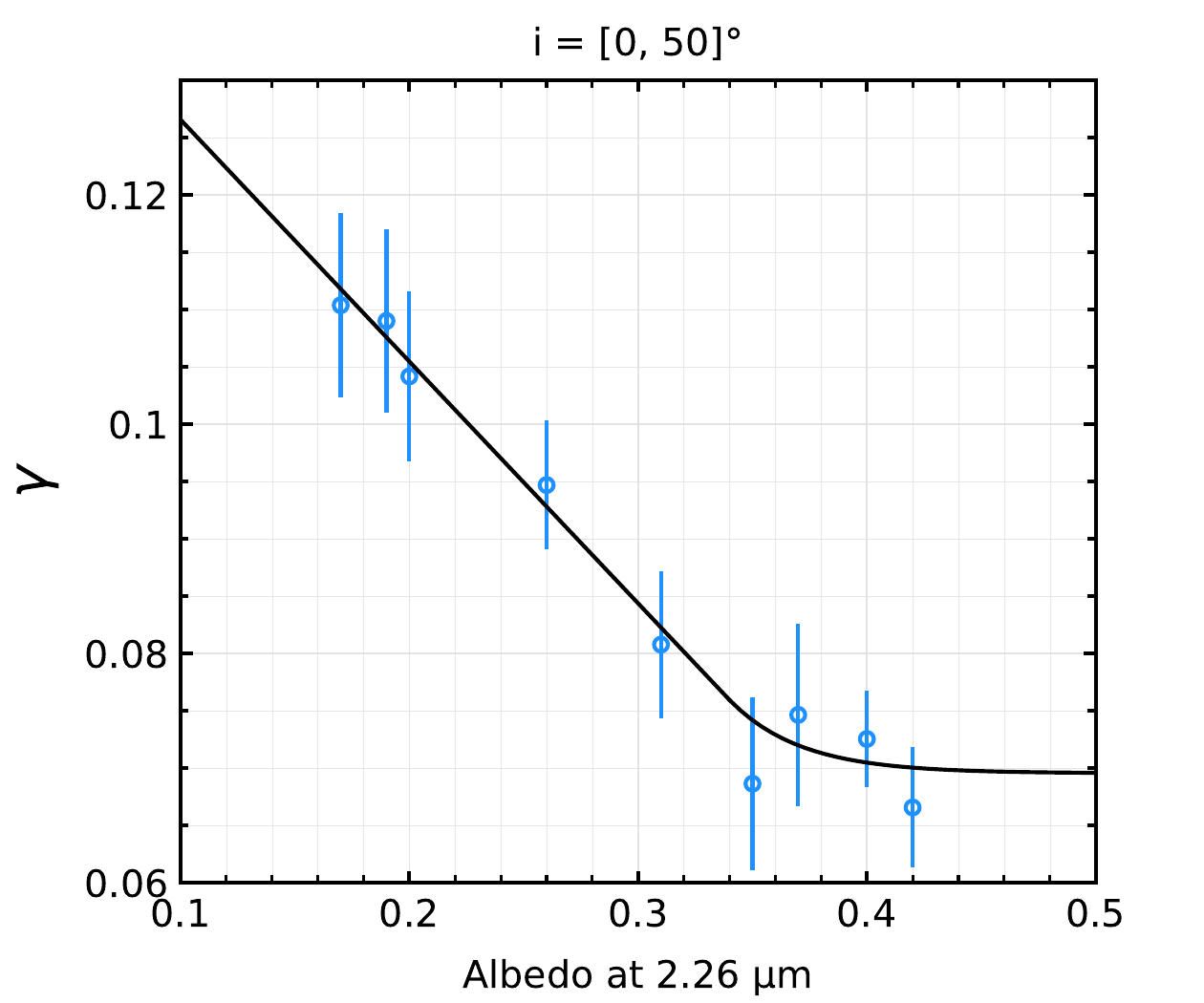}
	  \caption{Calibration of the $\gamma(\mbox{Alb})$ albedo dependence of $\Delta \tau_{CO_{2}} / \tau_{CO_{2},\ pred}$ (equation~\ref{eq_deltau_tau_MER_final_i0-50}). The slope of several fits similar to the two examples provided in Figure~\ref{graph_delta_tau_tau_MER_linfitAlb_i0-50} is shown as a function of albedo. The derived empirical function is detailed in equation~\ref{eq_director_factor_albedo_MER_i0-50} and Table~\ref{table_values_param_MER_calibration}.}\label{graph_coefdir_albedo_tauMER_i0-50}
\end{figure}

We first evaluate the relation between $\Delta \tau_{CO_{2}} / \tau_{CO_{2},\ pred}$ and $\tau_{MER}$ for low to moderate incidence angles ($\leq 50^\circ$) in Figure~\ref{graph_delta_tau_tau_MER_linfitAlb_i0-50}. We observe an overall increasing linear trend between both quantities (equation~\ref{eq_deltau_tau_MER_final_i0-50}), with some dispersion at low optical depths. We also notice that the slope $\gamma(\mbox{Alb})$ depends on surface albedo.

\begin{equation}
     \frac{\Delta \tau_{CO_2}}{\tau_{CO_{2},\ pred}} = \gamma(\mbox{Alb}) \ (\tau_{MER}-0.05)
    \label{eq_deltau_tau_MER_final_i0-50}
\end{equation}
We evaluate empirically $\gamma(\mbox{Alb})$ using data in Figure~\ref{graph_coefdir_albedo_tauMER_i0-50} (see legend for details). The fit parameters of resulting equation~\ref{eq_director_factor_albedo_MER_i0-50} are provided in Table~\ref{table_values_param_MER_calibration}.

\begin{equation}
    \gamma(\mbox{Alb}) = \left\{
    \begin{array}{ll}
        B \ \mbox{Alb} + C & \forall \mbox{Alb} \leq \mbox{Alb}_{cut}\\
        D \ e^{\kappa (\mbox{Alb}-\mbox{Alb}_{cut})} + E & \forall \mbox{Alb} \geq \mbox{Alb}_{cut}
    \end{array}
\right.
    \label{eq_director_factor_albedo_MER_i0-50}
\end{equation}

\begin{figure}[!h]
	\centering
		\includegraphics[width=0.5\textwidth]{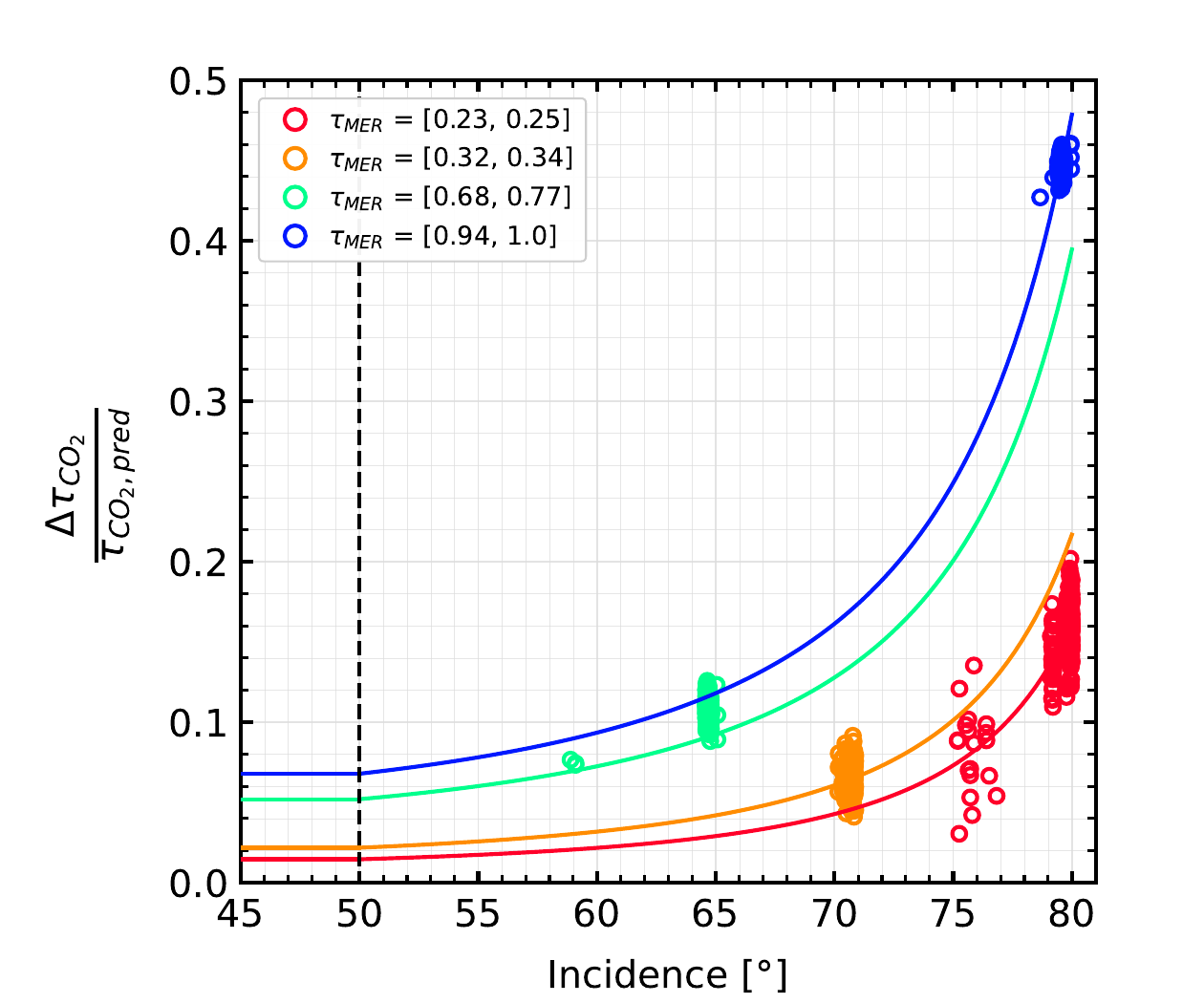}
	  \caption{$\Delta \tau_{CO_{2}} / \tau_{CO_{2},\ pred}$ as a function of incidence angles ($i > 45^\circ$) are shown for four different $\tau_{MER}$ ranges (colors). Albedos between $0.33$ and $0.34$ are only used. For $i < 50^\circ$ (symbolised by the vertical dashed line), the y-axis values are provided by equation~\ref{eq_deltau_tau_MER_final_i0-50}. For $i > 50^\circ$, we fit equation~\ref{eq_deltau_tau_MER_final_i50-80} (solid lines) on available data points (circles). Each fit differs by the exponent $\beta(\tau_{MER})$ value. }\label{graph_delta_tau_inci_MER_i0-80}
\end{figure}

\begin{figure}[!h]
	\centering
		\includegraphics[width=0.5\textwidth]{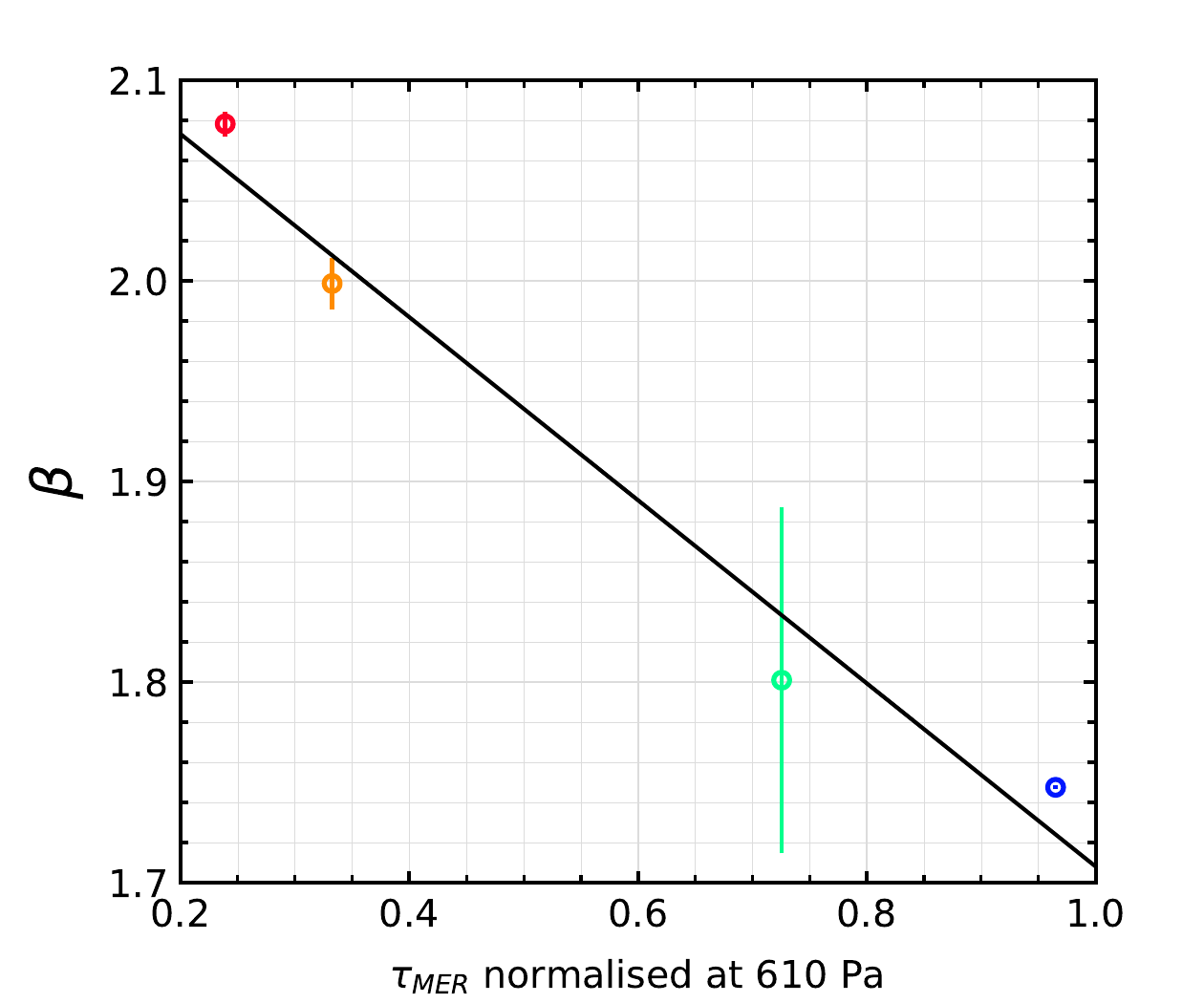}
	  \caption{Calibration of $\beta(\tau_{MER})$ (equation~\ref{eq_deltau_tau_MER_final_i50-80}). Each point is extracted from Figure~\ref{graph_delta_tau_inci_MER_i0-80} and used to constrain $\beta(\tau_{MER})$ function (black solid line, equation~\ref{eq_exponent_tauMER_i50-80}).}\label{graph_exponent_delta_tau_inci_MER_i0-80}
\end{figure}

At high incidence angles, most photons are scattered by atmospheric dust, which means that the CO$_2$ optical path is more sensitive to the presence of dust. Consequently, the relation between $\Delta \tau_{CO_{2}} / \tau_{CO_{2},\ pred}$ and $\tau_{MER}$ is modified for incidences greater than $50^\circ$. We evaluate this in Figure~\ref{graph_delta_tau_inci_MER_i0-80} where we show that for a given $\tau_{MER}$, $\Delta \tau_{CO_{2}} / \tau_{CO_{2},\ pred}$ increases according to the function defined in equation~\ref{eq_deltau_tau_MER_final_i50-80} ($i \geq 50^\circ$) where the exponent $\beta$ depends linearly on $\tau_{MER}$ (see Figure~\ref{graph_exponent_delta_tau_inci_MER_i0-80} , equation~\ref{eq_exponent_tauMER_i50-80} and Table~\ref{table_values_param_MER_calibration}).

\begin{equation}
     \frac{\Delta \tau_{CO_2}}{\tau_{CO_{2},\ pred}} = \gamma(\mbox{Alb}) \ (\tau_{MER}-0.05) \frac{1 + \Big (\frac{1}{\cos{i}}\Big )^{\beta(\tau_{MER})}}{1 + \Big (\frac{1}{\cos{50^\circ}}\Big )^{\beta(\tau_{MER})}}
    \label{eq_deltau_tau_MER_final_i50-80}
\end{equation}

\begin{equation}
    \beta(\tau_{MER}) = m \ \tau_{MER} + q
    \label{eq_exponent_tauMER_i50-80}
\end{equation}

Overall, we thus link $\Delta \tau_{CO_{2}} / \tau_{CO_{2},\ pred}$ and $\tau_{MER}$ through two equations: equation~\ref{eq_deltau_tau_MER_final_i0-50} for $i \leq 50^\circ$ and equation~\ref{eq_deltau_tau_MER_final_i50-80} for $i \geq 50^\circ$.

\begin{figure}[!h] 
	\centering
		\includegraphics[width=0.5\textwidth]{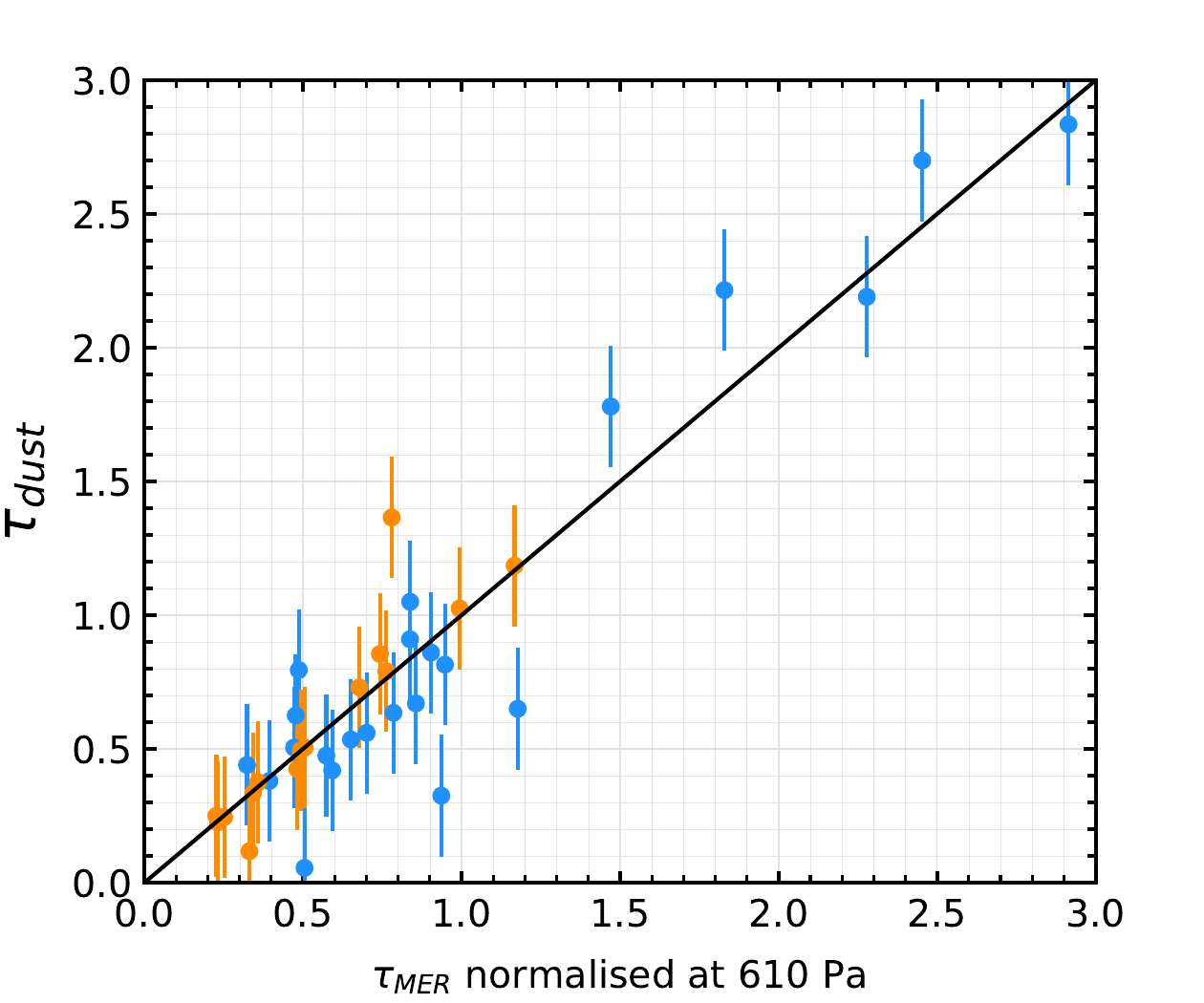}
	  \caption{Consistency of the MER calibration. This graph represents the median value of $\tau_{dust}$ for all OMEGA observations used for this calibration as a function of $\tau_{MER}$. Observations with a low-medium incidence angles ($\leq 50^\circ$) are symbolised by blue circles and those for higher angles by orange circles. The error bars represent the standard deviation between $\tau_{MER}$ and $\tau_{dust}$. The $y=x$ line is represented by a solid black line.}\label{graph_comp_tauOMEGA_tauMER}
\end{figure}

This calibration procedure makes it possible to convert a $\Delta \tau_{CO_{2}} / \tau_{CO_{2},\ pred}$ calculated with OMEGA data into an OMEGA dust optical depth $\tau_{dust}$ using a lookup table. This lookup table is populated by the previous equations, in which "$\tau_{MER}$" now corresponds to "$\tau_{dust}$". We illustrate the overall consistency of this approach  by comparing $\tau_{dust}$ to $\tau_{MER}$ in Figure~\ref{graph_comp_tauOMEGA_tauMER} for all available OMEGA/MER co-observations. We can notice that the point dispersion (from the $y=x$ line) does not depend on the dust optical depth. So, we compute the global standard deviation of the difference between $\tau_{MER}$ and $\tau_{dust}$: 0.2, which can be used as an uncertainty for $\tau_{dust}$ as represented in Figure~\ref{graph_comp_tauOMEGA_tauMER} as error bars. In the next section, we apply this calculation to the entire OMEGA dataset to derive NIR (0.9~µm) $\tau_{dust}$ and monitor atmospheric dust on Mars.

\begin{table}[!h]
\caption{Parameters values for equations~\ref{eq_deltau_tau_MER_final_i0-50}, ~\ref{eq_director_factor_albedo_MER_i0-50}, ~\ref{eq_deltau_tau_MER_final_i50-80}, and ~\ref{eq_exponent_tauMER_i50-80}.}\label{table_values_param_MER_calibration}
\begin{tabular*}{\tblwidth}{@{}CC@{}CC@{} }
\toprule
  & Parameters & Values\\ 
\midrule
 & $B$ & $-2.11 \times 10^{-1}$\\
 & $C$ & $1.48 \times 10^{-1}$\\
 & $D = B \ \mbox{Alb}_{cut} + C - E $ & $6.38 \times 10^{-3}$\\
 & $E$ & $6.95 \times 10^{-2}$\\
 & $\kappa$ & $3.21 \times 10^{1}$\\
 & $\mbox{Alb}_{cut}$ & $0.34$\\
 & $m$ & $-4.56 \times 10^{-1}$\\
 & $q$ & $2.16$\\
\bottomrule
\end{tabular*}
\end{table}

\section{Results and discussions}\label{section_results&discussions}

\subsection{Global seasonal maps}\label{subsection_seasonal_maps}
\begin{figure*}[!h]
	\centering
		\includegraphics[width=\textwidth]{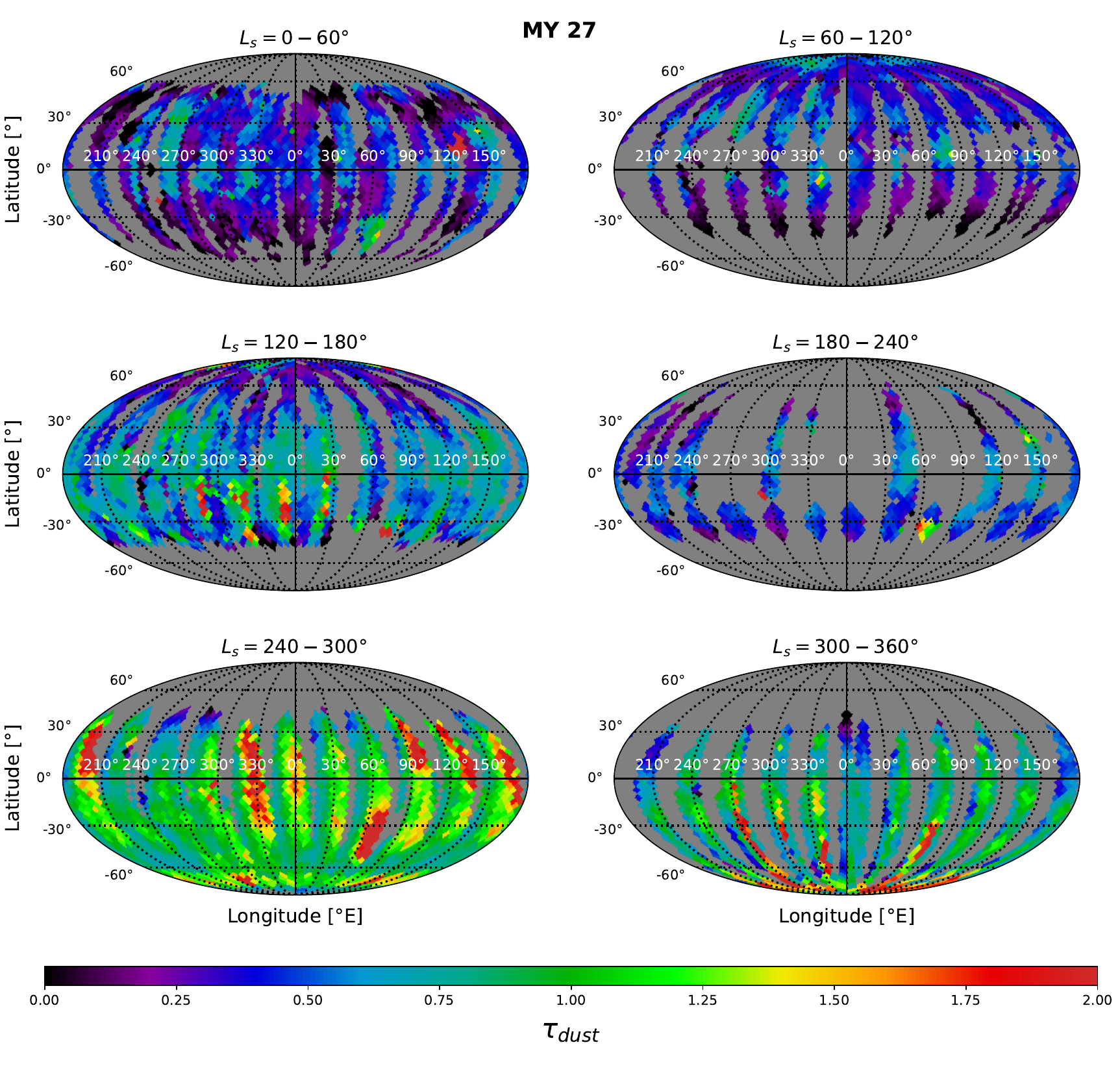}
	  \caption{Seasonal maps of dust optical depth computed with OMEGA data for MY~27. Each map corresponds to a range of 60$^\circ$ of solar longitude. A Mollweide projection is used. See section~\ref{subsection_seasonal_maps} for details. On the color bar, the black corresponds to $\tau_{dust}=0$ and the darkest red corresponds to $\tau_{dust} \geq 2$.}\label{figure_global_maps_MY27}
\end{figure*}

\begin{figure*}[!h]
	\centering
		\includegraphics[width=\textwidth]{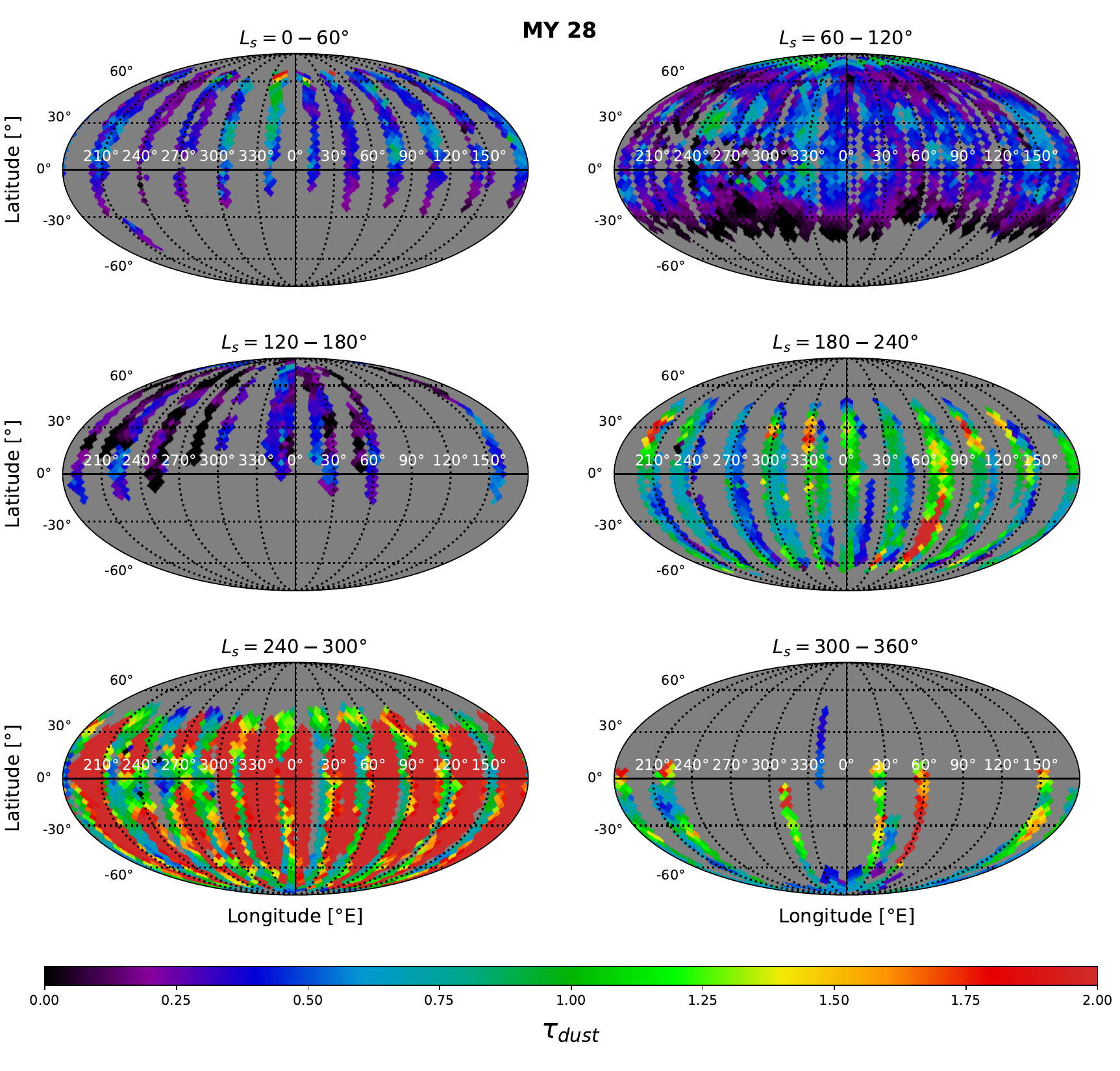}
	  \caption{Seasonal maps of dust optical depth computed with OMEGA data for MY~28.  Each map corresponds to a range of 60$^\circ$ of solar longitude. A Mollweide projection is used. See section~\ref{subsection_seasonal_maps} for details. On the color bar, the black corresponds to $\tau_{dust}=0$ and the darkest red corresponds to $\tau_{dust} \geq 2$.}\label{figure_global_maps_MY28}
\end{figure*}

\begin{figure*}[!h]
	\centering
		\includegraphics[width=\textwidth]{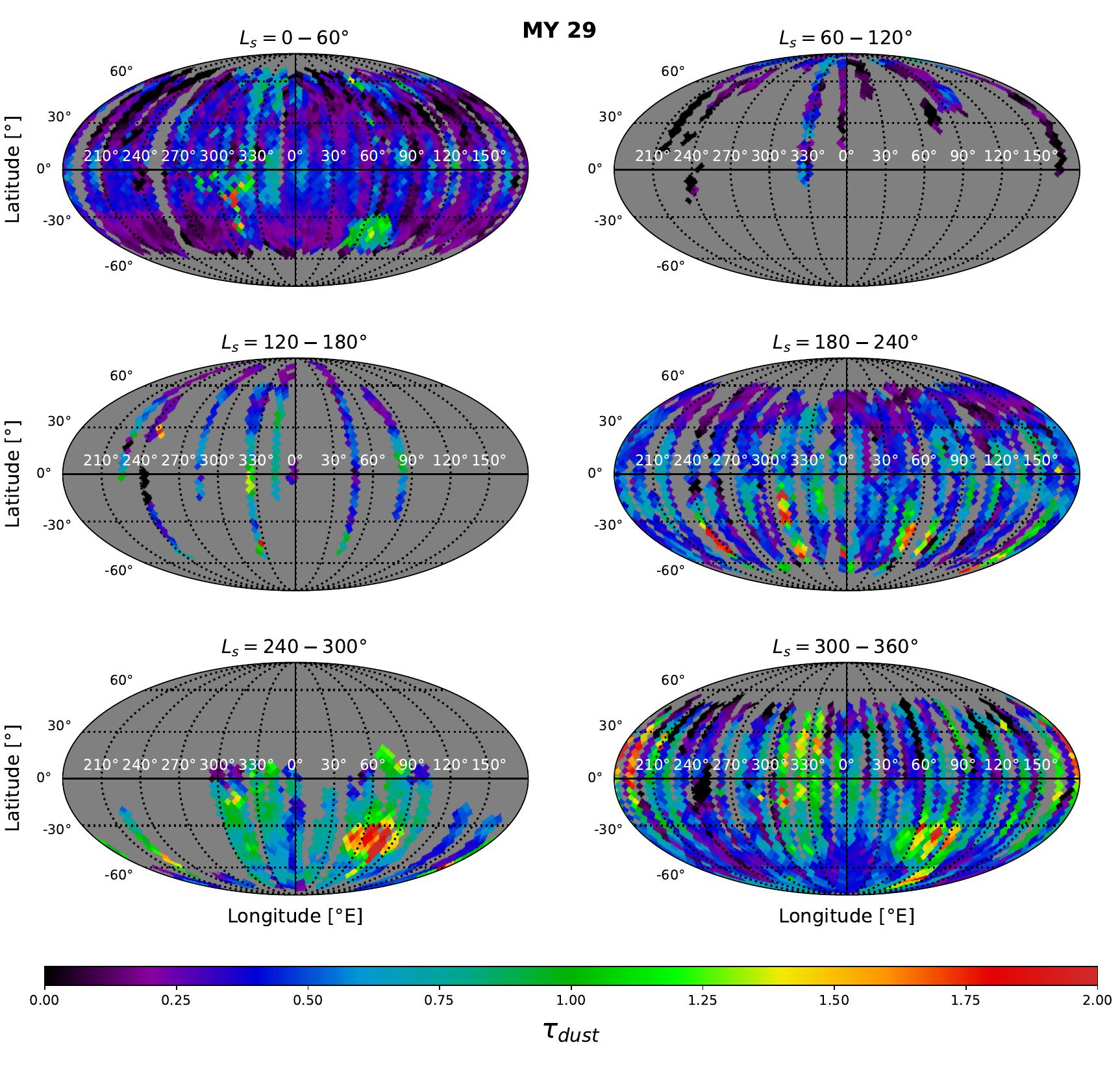}
	  \caption{Seasonal maps of dust optical depth computed with OMEGA data for MY~29. Each map corresponds to a range of 60$^\circ$ of solar longitude. A Mollweide projection is used. See section~\ref{subsection_seasonal_maps} for details. On the color bar, the black corresponds to $\tau_{dust}=0$ and the darkest red corresponds to $\tau_{dust} \geq 2$.}\label{figure_global_maps_MY29}
\end{figure*}

\begin{figure}[!h]
	\centering
		\includegraphics[width=0.495\textwidth]{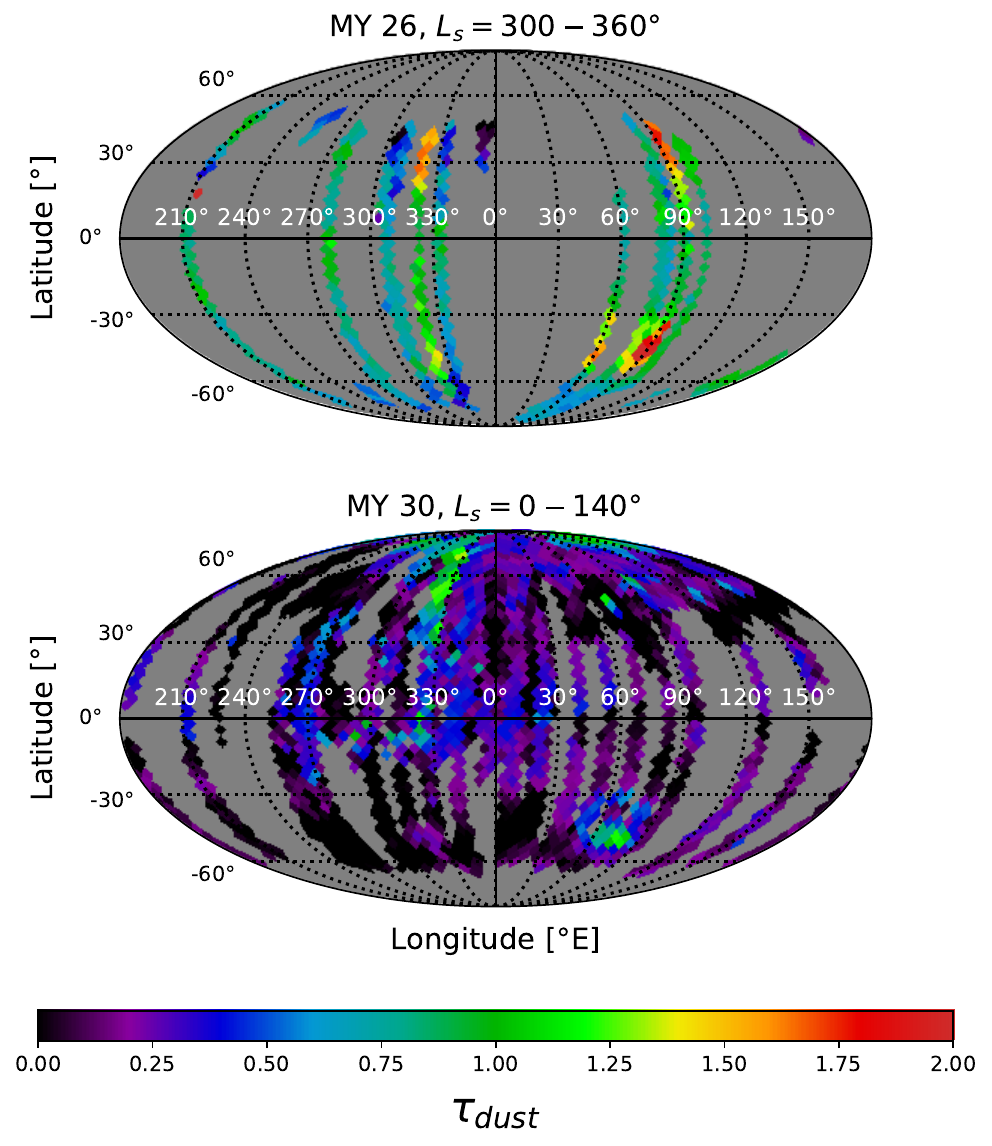}
	  \caption{Two additional maps of dust optical depth obtained during late MY~26 (top) and early MY~30 (bottom). MY~26 map starts from the arrival of Mars Express and MY~30 map goes up to the breakdown of OMEGA C-channel. These maps and the color bar are produced similarly as those presented in Figures~\ref{figure_global_maps_MY27}, \ref{figure_global_maps_MY28} and \ref{figure_global_maps_MY29}.}\label{figure_global_maps_MY2630}
\end{figure}

\begin{figure}[!h]
	\centering
		\includegraphics[width=0.495\textwidth]{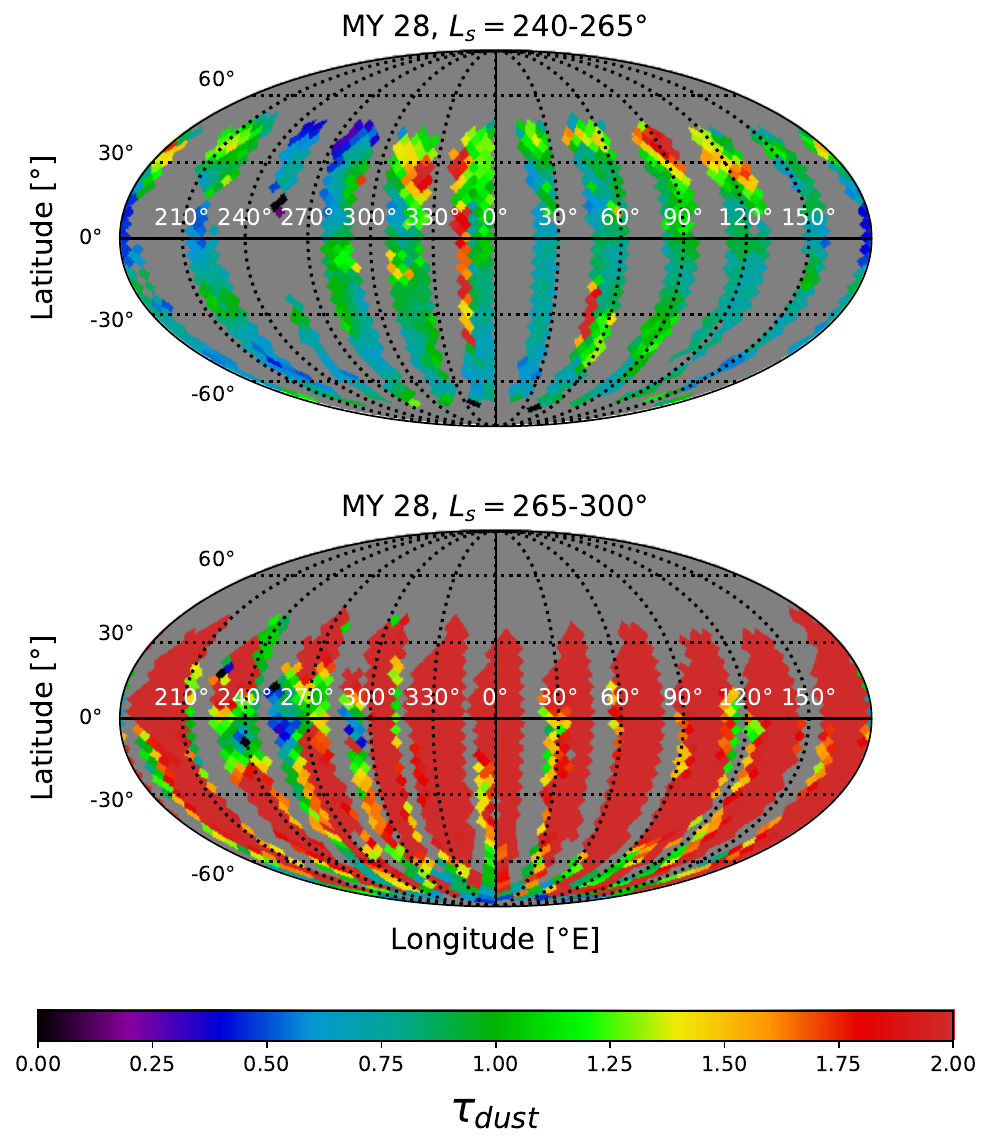}
	  \caption{Two sub-maps of dust optical depth derived from MY~28 $L_s=240-300^\circ$ map (bottom left panel of Figure~\ref{figure_global_maps_MY28}). On top: $L_s=240-265^\circ$ period that precedes the GDS,  and on bottom: $L_s=265-300^\circ$ period during the GDS. These maps and the color bar are produced similarly as those presented in Figures~\ref{figure_global_maps_MY27}, \ref{figure_global_maps_MY28} and \ref{figure_global_maps_MY29}.}\label{figure_global_maps_MY28_GDSperiod}
\end{figure}

We constructed global seasonal maps of dust optical depth using this OMEGA dataset. Each Martian year among MY~27, 28, and 29 is divided into six ranges of solar longitude $60^\circ$ wide (see Figures~\ref{figure_global_maps_MY27}, \ref{figure_global_maps_MY28} and \ref{figure_global_maps_MY29}). Two additional maps for late MY~26 and early MY~30 (beginning of Mars Express operations and end of the OMEGA C-channel respectively) were also constructed (Figure~\ref{figure_global_maps_MY2630}). We use the HEALPix pixelisation of the Martian sphere with an average spatial sampling of 5$^\circ$ in latitude and 6$^\circ$ in longitude, and we use a Mollweide projection (accuracy of proportions in area) to represent the maps. Each OMEGA observation cross several map pixels and a given map pixel can be covered by several OMEGA observations in a given solar longitude range. Consequently, the filling method of these maps is as follows: for each OMEGA observation and for a given map pixel, we compute the $\tau_{dust}$ median value of the OMEGA pixels located within the map pixel. If there are different OMEGA observations passing through the same map pixel, the maximum value between the median values of these observations is selected, so as to emphasis dust storm events (selecting the mean instead of the maximum only marginally changes these maps).

At global scale and first order, we observe the well-known seasonality of Martian dust: the global level of $\tau_{dust}$ is higher during the dust season ($L_s=180-300^\circ$) than in the clear atmosphere period ($L_s = 0-120^\circ$). At second order, we can see that the spatial distribution of atmospheric dust is not homogeneous regardless of the season: localised higher values are observed in all maps, even during the clear season, such as in MY~29 at $L_s=0-60^\circ$ (top left panel of Figure~\ref{figure_global_maps_MY29}) around $315^\circ$E and $20^\circ$S near Valles Marineris, where a dust storm is indeed present in the corresponding OMEGA observation.

We observe that some Martian areas have more atmospheric dust activity than others. For example, the Hellas bassin ($40-100^\circ$E, $20-60^\circ$S), a well-known dust source and sink \citep{cantor_martian_2001,szwast_surface_2006}, has high dust optical depths over most of the year. As discussed in the previous section, our procedure aims at estimating a dust optical depth normalised to a constant pressure level. Hence, these high values of optical depth over Hellas do correspond to a specific local increased in normalised opacity. Actually, we do not observe any systematic correlation between our (pressure normalised) dust optical depth and altitude. We can for example observe in Figure~\ref{figure_global_maps_MY29} that optical depth can be high over Hellas while being simultaneously low in some other low altitude areas (e.g., northern plains at $L_s=0-60^\circ$). We can also note that there is no optical depth decrease over high altitude areas in these maps, such as Tharsis ($\sim 245^\circ$E, $\sim 0^\circ$N), as could be seen if the dust optical depth was correlated with altitude.

These maps also highlight some major routes taken by dust storms according to previous studies, mainly from the northern hemisphere to the southern one \citep[see respectively their Figures~8 and 6]{battalio_mars_2021,wang_origin_2015}. Medium to high values of dust optical depth are indeed observed over these routes, such as for the Acidalia-Chryse ($\sim 325^\circ$E) channel, or Arcadia-Cimmeria/Sirenum ($\sim 180^\circ$E) route (e.g., see MY~29 $L_s=300-360^\circ$ map in Figure~\ref{figure_global_maps_MY29}).

The OMEGA dataset covers the MY~28 GDS that occurred between $L_s \sim 265^\circ$ and $310^\circ$ (Figure~\ref{figure_global_maps_MY28}) which is detailed in two sub-maps in Figure~\ref{figure_global_maps_MY28_GDSperiod}. We observe high dust optical depths from Chryse to Noachis ($\sim 345^\circ$E) just before the beginning of the GDS in the $L_s=240-265^\circ$ range (more precisely at $L_s=264.4^\circ$). This area corresponds to two dust storms also reported in \cite{wang_origin_2015} before the GDS onset. GDS initiation probably occurred slightly later at the east of Hellas ($\sim 50^\circ$E) according to \cite{wang_origin_2015}, where and when we also observe high dust optical depths in an OMEGA observation obtained at $L_s=264.9^\circ$ (Figure~\ref{figure_global_maps_MY28_GDSperiod}, top panel).

\subsection{Comparison with previous thermal IR studies} \label{subsection_TIR_comparisons}
\begin{figure*}[!h]
	\centering
		\includegraphics[width=0.8\textwidth]{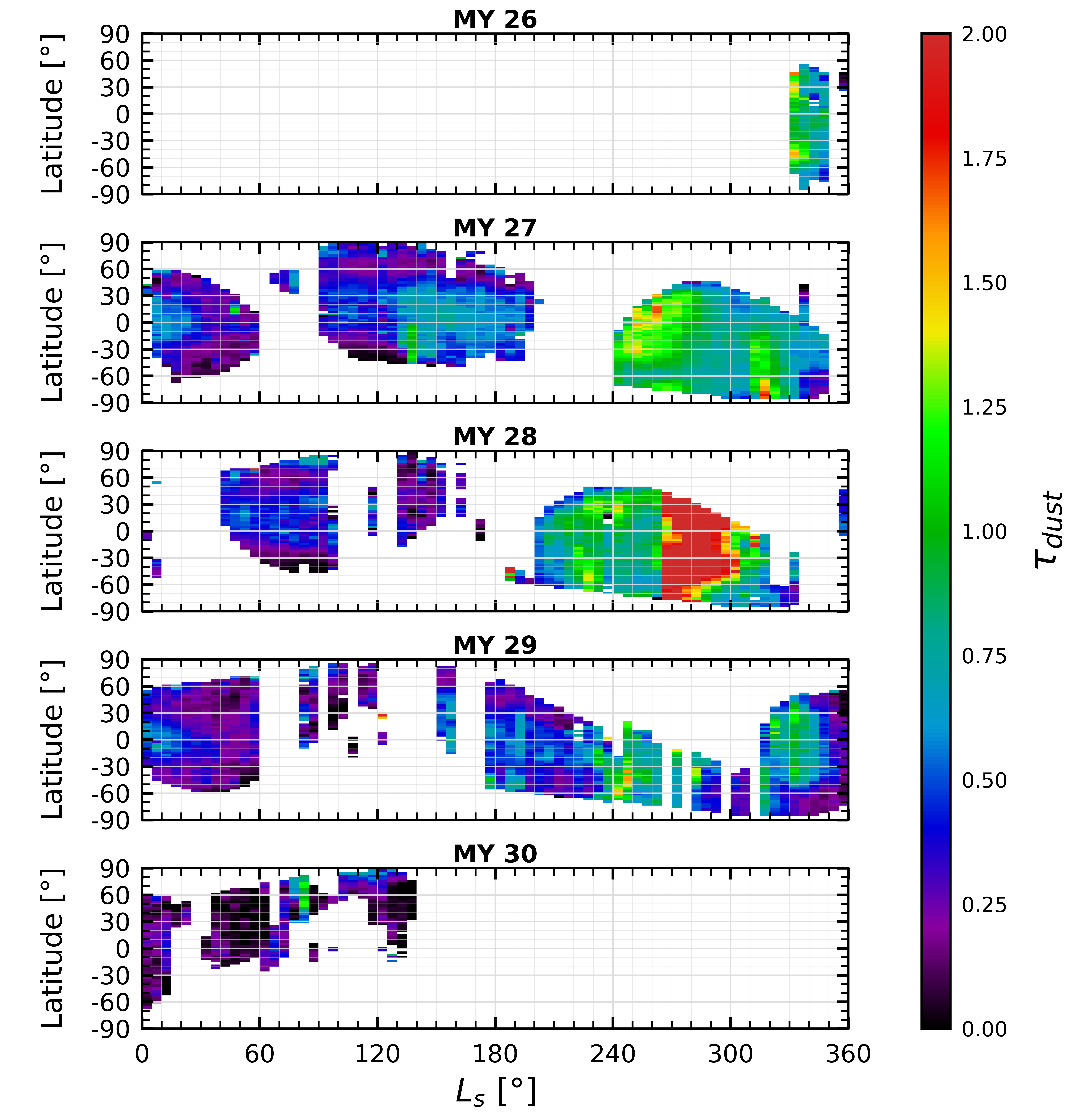}
	  \caption{Overview of atmospheric dust distribution for the three full Martian years (MY~27, 28, 29) and the two partial years corresponding to the beginning (MY~26) and the ending (MY~30) of OMEGA C-channel operations. Zonal means of dust optical depth maps as a function of latitude and season ($L_s$) are shown. For each solar longitude range of $5^\circ$, we computed the average of $\tau_{dust}$ for all longitudes and for latitude bin about $3.2^\circ$ width.}\label{figure_lat/ls_diagrams}
\end{figure*}

From the OMEGA dataset we can also produce latitude/solar longitude diagrams of dust optical depth for MY~26, 27, 28, 29 and 30 (see Figure~\ref{figure_lat/ls_diagrams}). At first order, observed relative variations with season and latitude are in good agreement with those derived from Thermal-InfraRed (TIR hereinafter) data \citep[see their Figure~16]{montabone_eight-year_2015}. At second order, we can notice a few differences. For example, $L_s=305-335^\circ$ storms (named "C-storms" in \cite{kass_interannual_2016}) have lower values of optical depth compared to $L_s=240-300^\circ$ storms in TIR retrievals \citep{montabone_eight-year_2015}, while both periods have similar NIR optical depth levels (see Figure~\ref{figure_lat/ls_diagrams}).

\begin{figure}[!h]
	\centering
		\includegraphics[width=0.49\textwidth]{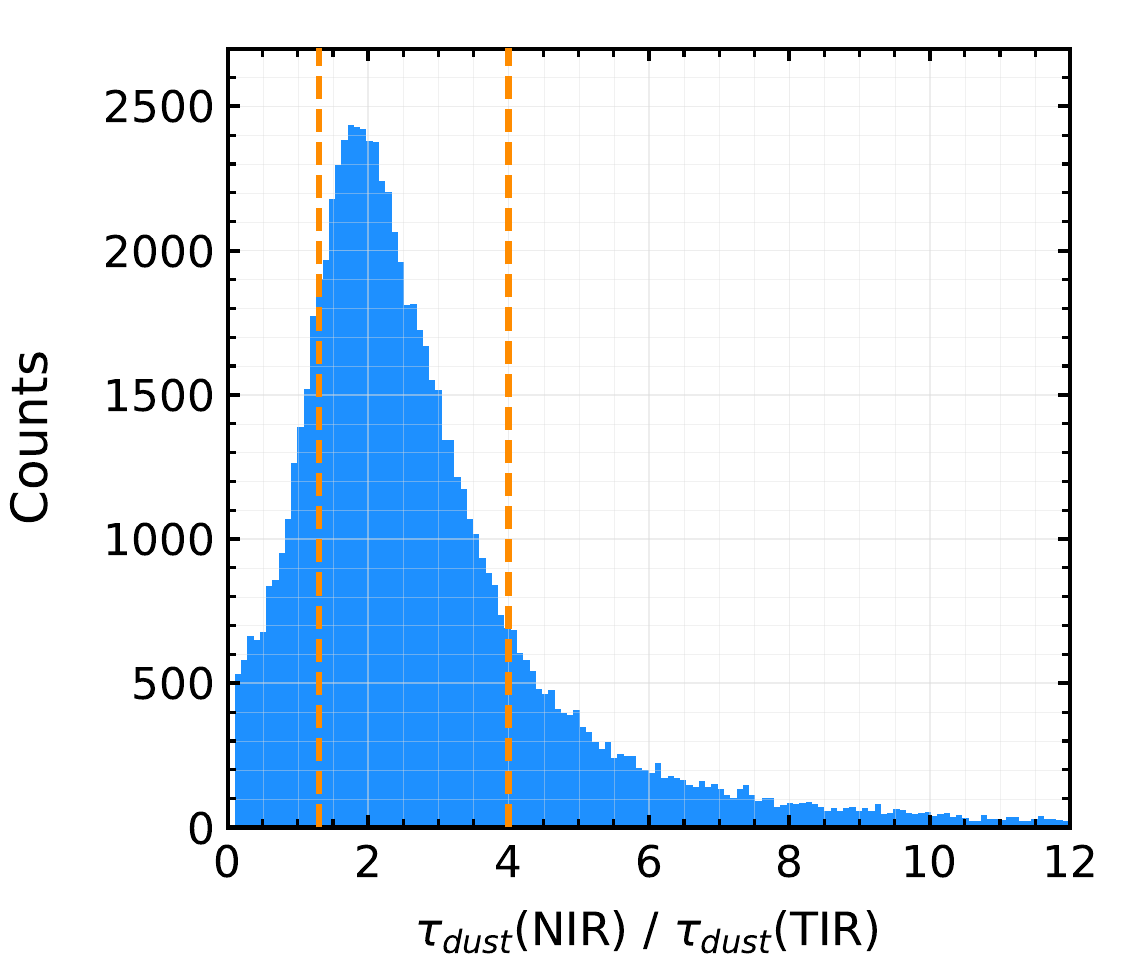}
	  \caption{Histogram of the dust extinction optical depth ratio between NIR (OMEGA, this study) and TIR \citep{montabone_eight-year_2015}. The distribution peaks at 1.8. The vertical dashed orange lines represent ratio values 4 and 1.35 that correspond to dust particle size $r_{eff}=0.7$ and $2.1$ respectively (for an effective variance of 0.3) according to \cite{wolff_constraints_2006}.}\label{figure_histo_NIR/TIR}
\end{figure}

These differences can be studied by computing the NIR/TIR dust optical depth ratio. Such variations of this ratio have implications for general circulation models, as these models rely on radiative budget calculations in both the solar and the TIR range \citep{forget_improved_1999}. Hence, the knowledge of the dust optical depth NIR/TIR ratio is required, in addition to the amount of dust, to perform accurate predictions. So, we have calculated the ratio between our dust extinction optical depth values derived in NIR (OMEGA) and TIR 9~µm extinction Column Dust Optical Depth (CDOD) normalised at 610~Pa provided by \cite{montabone_eight-year_2015} on the basis of contemporary TES/MGS, THEMIS/MO and MCS/MRO data (see their Figure~1). 
The resulting histogram is shown in Figure~\ref{figure_histo_NIR/TIR}. The ratio distribution peaks at 1.8, which is in agreement with the values (between 1.25 and 5) computed locally with MER data in \cite{lemmon_dust_2015} using Pancam in NIR and Mini-TES in TIR (see their Figure~10). This average ratio is also consistent with \cite{montabone_eight-year_2015} who consider a 2.6 value for the NIR extinction to TIR absorption ratio. Indeed, this corresponds to an extinction/extinction ratio of 2 using a 1.3 conversion factor between absorption and extinction at 9.3~µm \citep{wolff_constraints_2006}.
We can notice that the distribution is very large with a 1.5 "equivalent standard deviation" (defined by the half difference of the boundaries of the interval that contains 68\% of the values). Knowing that the NIR/TIR ratio depends on the dust particle size \citep[see their Figure~13]{wolff_constraints_2006}, we compare our distribution with expected variations caused by particle size in Figure~\ref{figure_histo_NIR/TIR}. Size variations as a function of place and season probably explain part of the distribution width, apart from very high or very low ratio values. These latter could correspond to specific events observed by one of the two instruments only, or this may result from retrieval biases. Actually, biases specific to each method have already been identified. For example, the impact of a given amount of atmospheric dust on the CO$_2$ optical depth will depends on dust vertical distribution (e.g., well-mixed or not), while we do not include this level of refinement in our modelling.

\begin{figure}[!h]
	\centering
		\includegraphics[width=0.49\textwidth]{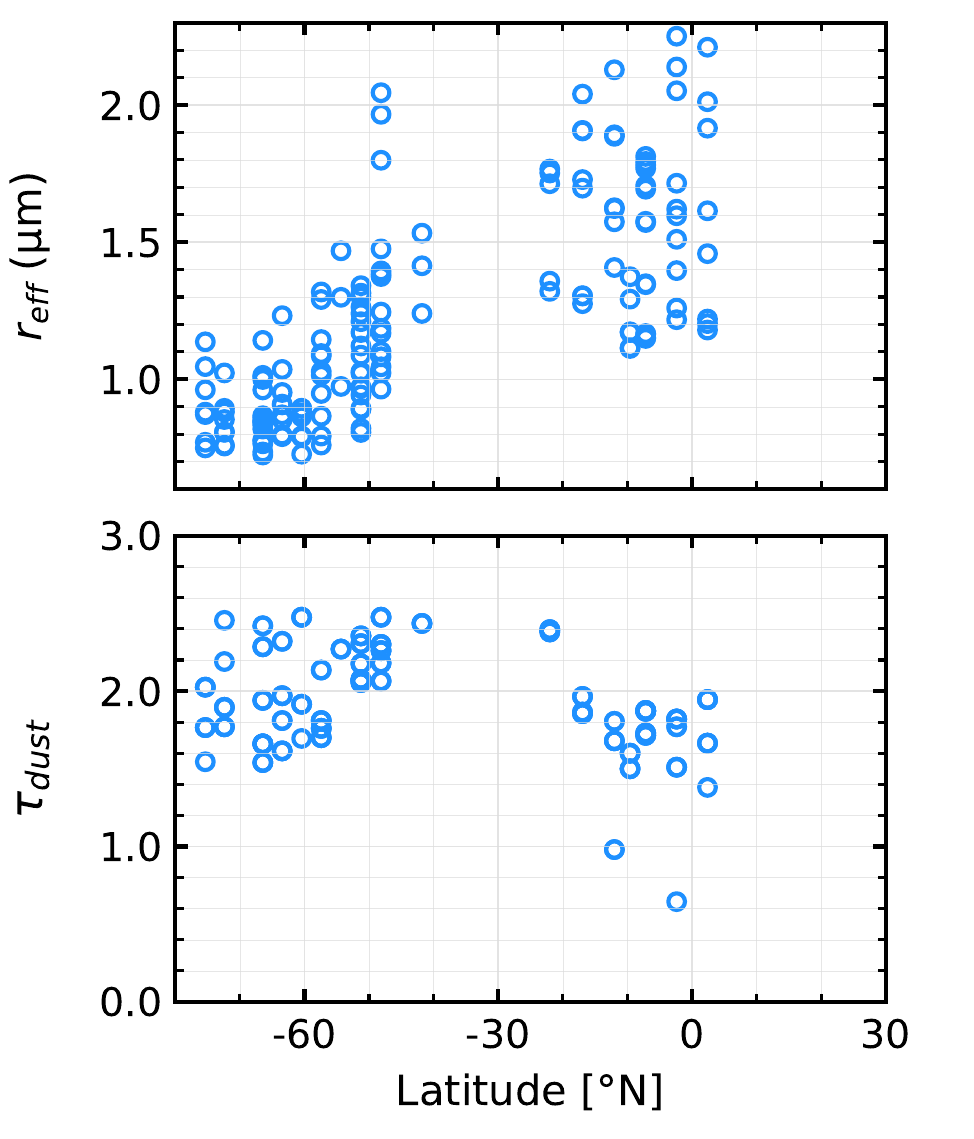}
	  \caption{Variations with latitude of (top) dust effective radius and (bottom) dust optical depth during MY~28 for $L_s=[271, \ 275]^\circ$. All longitudes are considered (see section~\ref{subsection_TIR_comparisons} for details).}\label{figure_reff_lat}
\end{figure}

\begin{figure}[!h]
	\centering
		\includegraphics[width=0.49\textwidth]{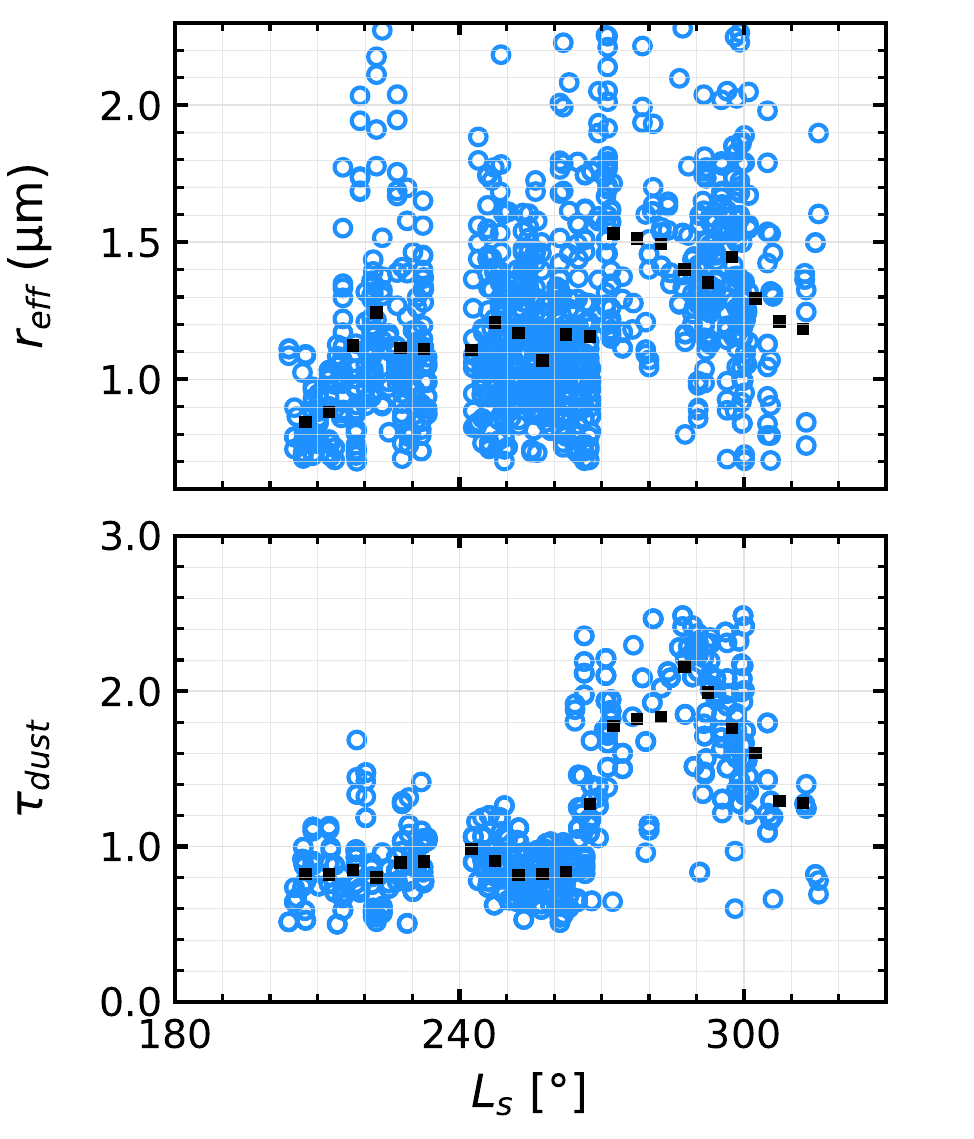}
	  \caption{Variations with solar longitude of (top) dust effective radius and (bottom) dust optical depth during MY~28. Longitudes in the $[-10, \ 10]^\circ$N range are considered (each blue circle corresponds to a map pixel). Black squares represent the mean values for $5^\circ$ $L_s$ range width (see section~\ref{subsection_TIR_comparisons} for details).}\label{figure_reff}
\end{figure}

\subsection{Particle size variations during the MY~28 GDS}

We used results presented in Figure~13 of \cite{wolff_constraints_2006} to convert the dust extinction optical depth NIR/TIR ratio to $r_{eff}$ (considering an effective variance of 0.3). We restrict this conversion to OMEGA $\tau_{dust}=[0.5, \ 2.5]$ and NIR/TIR ratio values between 1 and 4. We show two examples of particle size variations observed with this method during the 2007/MY~28 GDS in Figures~\ref{figure_reff_lat} and \ref{figure_reff}.

During the peak of the GDS (maximum latitudinal extension, $L_s=[271, \ 275]^\circ$), we observe that particle size decreases with latitude from the equator towards the south pole, while optical depth is high and nearly constant with latitude (Figure~\ref{figure_reff_lat}). Latitudinal variations of $r_{eff}$ have already been observed in \cite{wolff_constraints_2003} for the 2001A GDS (see their Figures~15 and 16). This observation could be explained by the greater ability of small particles to be transported (e.g., by the Hadley cell circulation) far from the GDS initiation area.

We also note an overall increase and then decrease of the particle size that correlates in time respectively with the onset and decay of the 2007 GDS (Figure~\ref{figure_reff}). Such an increase in the effective dust particle size is expected during dust storms, due to convective movements and vertical mixing overcoming gravitational sedimentation of dust particles (e.g., \cite{haberle_GDS_effects_circulation_1982}, \cite{kahre_sizes_2008}, \cite{spiga_rocket_2013}) and dust coagulation processes \citep{bertrand_coagulation_2022}. For instance, the modelling of the 2018 GDS revealed strong updrafts associated with plumes of dust and an increase in particle size \citep{bertrand_simulation_2020}, which was confirmed by observations ("dust towers", \cite{heavens_convection_2019}).
The median value of the effective radius varies from $1.1$~µm before/after the GDS to $1.5$~µm during the GDS. This average trend is in agreement with retrievals reported during the same GDS with another method by \cite{vincendon_yearly_2009} (see their Figure~15). Note that maximum values reach $2.3$~µm (Figure~\ref{figure_reff}), which is the maximum possible value of our retrieval method (see Figure~13 of \cite{wolff_constraints_2006}). The existence of higher values up to 3-4~µm reported for other GDS \citep{wolff_constraints_2003,lemmon_large_2019} is not evaluated here.

\subsection{Search for MY~28 GDS precursory signs} \label{subsection_GDS}

We can notice that our $\tau_{dust}$ map obtained in MY~27 in the $L_s=240-300^\circ$ range (storm season) (Figure~\ref{figure_global_maps_MY27}) is similar to the one obtained by MGS/TES  in MY~24 in the same $L_s$ range \citep[see their Figure~8]{wang_origin_2015}. As both MY~27 and 24 are years that precede a GDS (2007 and 2001 respectively), we have looked for differences between the year before and after the 2007 GDS, so as to see if the year preceding the storm shows precursory signs.

The different spatial coverage of our global maps and the large time range of these maps can introduce some biases. Hence, it is necessary to use both global maps (Figures~\ref{figure_global_maps_MY27}, \ref{figure_global_maps_MY28}, \ref{figure_global_maps_MY29}, and \ref{figure_global_maps_MY2630}) and latitude/$L_s$ diagrams (Figure~\ref{figure_lat/ls_diagrams}) to extract trends. We observed that MY~27 (before the GDS) and MY~29 (after the GDS) are comparable in term of dust optical depth at a year scale, with few differences such as a regional dust storm in MY~27. This is in agreement with local MER measurements from the surface \citep[see their Figure~7]{lemmon_dust_2015} and global TIR measurements from the orbit \citep[see their Figure~16]{montabone_eight-year_2015}. Thus, we do not observe obvious significant distinctive signs the year before the MY~28 GDS.

The combination of these three datasets however suggests some differences not the year, but the season preceding the GDS. If we compare this pre-GDS season (from $L_s \sim 150^\circ$ to $260^\circ$) between MY~27, 28 and 29, we observe lower optical depths in MY~28, in particular between $L_s \sim 240^\circ$ and $260^\circ$, as previously noticed in Meridiani by \cite{lemmon_dust_2015}. The 2007 GDS is a late GDS contrary to the 2001 and 2018 one: such lower activity in the early storm season could thus be a precursory sign of a late storm season GDS.

\subsection{Link with RSL} \label{subsection_RSL}

\begin{figure*}[!h]
	\centering
		\includegraphics[width=0.7\textwidth]{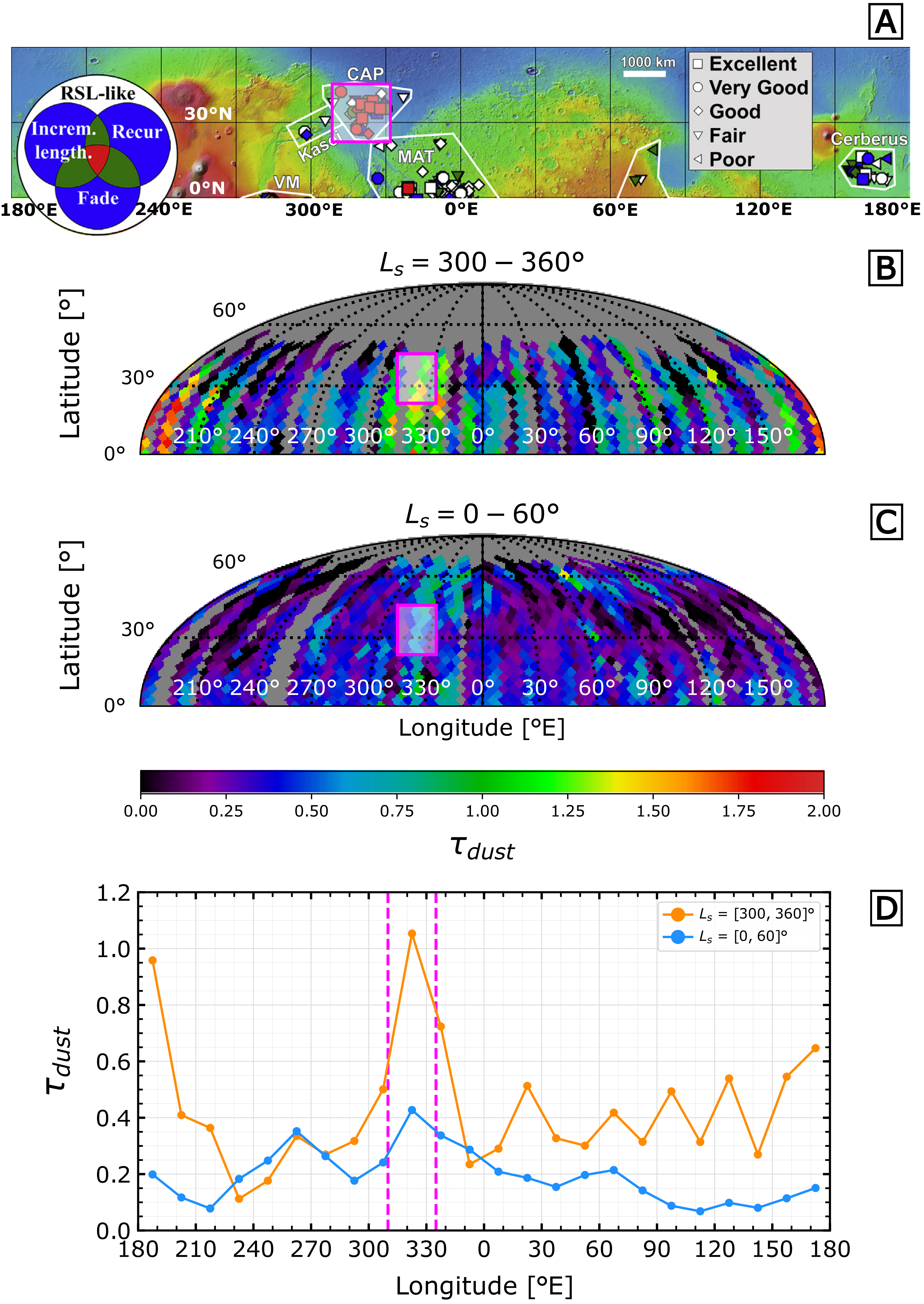}
	  \caption{Illustration of the time and spatial correlation between atmospheric dust and RSL in the northern hemisphere. [A] is a modified version of Figure~2.1 of \cite{stillman_chapter_2018} which represents the distribution of RSL in different sites on a topographic map (here for the northern hemisphere), with a specific legend (color and shape) to characterize the RSL detections. [B] and [C] are MY~29 seasonal global maps (northern hemisphere) of dust optical depth. The pink rectangle corresponds to the main area of confirmed RSL (red symbols) in the northern hemisphere from panel~[A]. [D] longitudinal profile of dust optical depth from [B] (orange) and [C] (blue) maps, for a $25^\circ - 45^\circ$N latitude ring that includes most northern hemisphere confirmed RSL (pink rectangle). The longitudes of RSL are indicated with pink dotted lines.}\label{figure_RSL}
\end{figure*}

Recurring Slope Lineae (hereinafter RSL) may form when and where unstable dust has been recently deposited on the surface, and when and where winds are strong enough to remove dust or initiate dust removal on a slope \citep{vincendon_observational_2019}. In the northern hemisphere, we can see in the panel~A of Figure~\ref{figure_RSL} that confirmed RSL (red markers) have been predominantly reported near Chryse-Acidalia Planitia \citep{stillman_observations_2016, stillman_chapter_2018}. These RSL take place inside the Acidalia-Chryse dust storm travel route discussed previously (see section~\ref{subsection_seasonal_maps}), which is also a low albedo area (like most RSL sites, \cite{mcewen_seasonal_2011}). These two characteristics (low albedo and high atmospheric dust activity) show that Chryse-Acidalia is an area where winds are strong enough to remove dust from the surface \citep{thomas_mesoscale_2003}, and to initiate dust storms or to transport atmospheric dust. Additionally, we observe that this RSL area has higher dust optical depth values than its surrounding during two periods of major RSL formation (i.e., when dust is expected to be removed from the surface): $L_s=300 - 360^\circ$ and $L_s=0 - 60^\circ$ (see respectively panels~B and C of Figure~\ref{figure_RSL}). Actually, at mid-latitudes, the longitudes of Chryse-Acidalia RSL are the only ones that correspond to a local maximum of atmospheric dust optical depth for both $L_s=300 - 360^\circ$ and $L_s=0 - 60^\circ$ (see panel~D of Figure~\ref{figure_RSL}). Thus, these northern hemisphere RSL are correlated with atmospheric dust in time and space.

\section{Summary and conclusions}\label{section_conclusion}
\begin{itemize}
\item We have developed and applied a new method to detect atmospheric dust in the Mars Express/OMEGA Near-InfraRed (NIR) dataset. Atmospheric dust reduces the strength of the 2~µm CO$_2$ gas absorption band: comparing the observed bands with model predictions for clear atmosphere makes it possible to detect atmospheric dust.

\item We then analysed co-observations from the orbit by OMEGA and from the surface by the Mars Exploration Rovers to calibrate OMEGA dust detections and to convert our measurements at 2~µm to 0.9~µm "NIR" dust optical depths.

\item We applied this method to the OMEGA dataset to derive seasonal global maps of dust optical depth from late MY~26 to mid MY~30 (Figure~\ref{figure_global_maps_MY27}, \ref{figure_global_maps_MY28}, \ref{figure_global_maps_MY29} and \ref{figure_global_maps_MY2630}). Main well-known features of the dust cycle, such as dust storm travel routes (Acidalia-Chryse $\sim 325^\circ$E, Arcadia-Cimmeria/Sirenum $\sim 180^\circ$E) or major dust sources/sinks (e.g., Hellas) are recognisable.

\item We did not identify major significant differences between the year preceding and the year following the Global Dust Storm (GDS). However, we note that the beginning of the dust season prior to the onset of the MY~28 GDS was less intense in MY~28 compare to the same period of MY~27 and MY~29.

\item We compared our NIR (0.9~µm) dust optical depth with TIR (9~µm) values derived in previous studies: the NIR/TIR extinction optical depth ratio is 1.8 on average (Figure~\ref{figure_histo_NIR/TIR}) which is in agreement with previous results.

\item We have converted this optical depth ratio to particle size. We observed latitudinal and seasonal variations of dust mean particle size linked to the MY~28 GDS.

\item We observed that Recurring Slope Lineae (RSL) location and timing in the northern hemisphere correlate with atmospheric dust activity (Figure~\ref{figure_RSL}). This correlation suggests that the RSL activity is linked to surface dust rising and/or atmospheric transport phenomena.

\item Finally, we can notice that this method could be adapted to other orbital NIR imaging spectrometer datasets such as CRISM/MRO or MIRS/MMX.

\end{itemize}

\section*{Acknowledgments}
The OMEGA/Mex data are freely available on the ESA PSA at \url{https://archives.esac.esa.int/psa/#!Table%20View/OMEGA=instrument}.

The authors thank the reviewers for their pertinent and helpful comments that really improve the quality of this paper. 
The first author also wants to thank A. Stcherbinine for having developed such a useful Python tool (OMEGA-Py).

The authors have a thought for Brigitte Gondet who passed away this year and who was an important member of the Mars Express mission, working among other things to improve the OMEGA dataset from the beginning of the mission until now. Her person and expertise will be missed. Merci Brigitte!

\printcredits
\bibliographystyle{cas-model2-names}

\bibliography{cas-dc-template}

\end{document}